**DFT+U+J with linear response parameters predicts non-magnetic oxide band gaps with hybrid-functional accuracy**

*D. S. Lambert[a] and D. D. O'Regan[a*]*

a. School of Physics, SFI AMBER Centre and CRANN Institute, Trinity College Dublin, The University of Dublin, Ireland

**Abstract:**

First-principles Hubbard-corrected approximate density-functional theory (DFT+U) is a low-cost, potentially high throughput method of simulating materials, but it has been hampered by empiricism and inconsistent band-gap correction in transition-metal oxides. DFT+U property prediction of non-magnetic systems such as $d^0$ and $d^{10}$ transition-metal oxides is typically faced with excessively large calculated Hubbard U values, and with difficulty in obtaining acceptable band-gaps and lattice volumes. Meanwhile, Hund's exchange coupling J is an important but often neglected component of DFT+U, and the J parameter has proven challenging to directly calculate by means of linear response. In this work, we provide a revised formula for computing Hund's J using established self-consistent field DFT+U codes. For non-magnetic systems, we introduce a non-approximate technique for calculating U and J simultaneously in such codes, at no additional cost. Using unmodified Quantum ESPRESSO, we assess the resulting values using two different DFT+U functionals incorporating J, namely the widely used DFT+(U-J) and the readily available DFT+U+J. We assess a test set comprising $TiO_2$, $ZrO_2$, $HfO_2$, $Cu_2O$ and ZnO, and apply the corrections both to metal and oxygen centered pseudoatomic subspaces. Starting from the PBE functional, we find that DFT+(U-J) is significantly out-performed in band-gap accuracy by DFT+U+J, the mean-absolute band-gap error of which matches that of the hybrid functional HSE06. ZnO, a long-standing challenge case for DFT+U, is addressed by means of Zn 4s instead of Zn 3d correction, whereupon the first-principles DFT+U+J band-gap error falls to half of that reported for HSE06, yet remains larger than for PBE0.

# 1. Introduction

The electronic bandgap of a material is a property of high importance for semiconductor and photovoltaic applications. However, attempts to computationally model the bandgap using approximate Density Functional Theory (DFT) are consistently inaccurate[1]. This inaccuracy arises because the functionals used, such as the Local Density Approximation (LDA) and the Generalized Gradient Approximation (GGA), give rise to self-interaction error (SIE), among other systemic errors, leading to an underestimation of the band-gap. This can often be partially ameliorated through the use of hybrid functionals such as HSE06[2] and PBE0[3,4]. These functionals are often much more accurate at predicting band-gaps than purely local or semi-local DFT approximations, but can be much more computationally expensive, depending on the implementation and simulation size. This has motivated the search for, alongside others discussed in the recent review of Ref. [5], spatially localized corrections which act upon cost-effective semi-local DFT and which allow for accurate band-gap modelling without the need for hybrid-level computational costs[6-9].

Perhaps the most common correction applied to conventional DFT functionals is the Hubbard model inspired method now widely referred to as DFT+U[10-12], which in its simplest form effectively adds an energy penalty for partial occupation of pre-defined sets of localized orbitals. This method requires the choice of a value for the 'U parameter' to set the magnitude of the correction. The U value is often tuned to try and match experimental values, sometimes even those of quantities that are not ground-state DFT-accessible observables. This diminishes the first-principles nature of the DFT methodology and does not allow for the accurate prediction of gaps for less well-characterized or only theoretically predicted materials, where the band-gap values are necessarily not well known. In order to avoid the use of ad-hoc or empirical U values, several methods have been proposed for calculating the U value from first principles, such as the well-known finite-difference linear response method [13-15]. There, a U value is calculated in terms of the response of the occupancies of the aforementioned localized orbitals to an applied perturbation.

The use of DFT+U has faced many difficulties in properly modelling transition-metal oxide (TMO) band-gaps. The corrective terms of DFT+U are most commonly applied to pre-defined pseudo-atomic d orbitals of the metal ion(s) only, which can still result in a significant bandgap underestimation in $d^0$ or $d^{10}$ systems [16,17]. The application of DFT+U to metal orbitals alone also can result in a distortion of the geometric structure of the material[18-20]. It has become more common in recent years to apply U parameters to both the metal d-orbitals and the O 2p orbitals[21-27], however the U corrections calculated with linear response methods applied to all atoms have then tended to result in a band-gap overestimation[27]. In addition, attempts to calculate the U with linear response on non-magnetic oxides with nearly full or empty d orbitals such as ZnO have tended to fail, giving unreasonably high U values[23,28] or resulting in numerical instability in the calculations[29,30].

In addition to the Hubbard U correction, Hund's J correction has long been incorporated to better account for the effect of intra-atomic exchange. However, with Hund's J in the present work we additionally target spin-flip exchange, a beyond-Hartree-Fock effect [31] and therefore a correction for what can be considered a genuine correlation (multi-reference) effect [32]. We refer the reader to Ref. [33] for a discussion of how different exchange terms relate to Hund's 1st and 2nd rules. In this work, specifically, two different corrective functionals incorporating U and J parameters were evaluated. The first is the well-known functional introduced by Dudarev *et al.*[34], which is a rotationally invariant simplification of previous formulations from Ref. [11] and Ref. [35]. This technique involves combining U and J into an effective parameter $U_{eff}$ = U-J. For the sake of brevity this functional will be referred to as DFT+(U-J) in this study, with the brackets emphasizing that the calculated U and J values are combined into one parameter for a single energy correction term. The Dudarev functional was chosen here for its widespread availability and popularity in both condensed matter physics and solid state chemistry.

The second functional, which is newer and the main focus of this study, originated with Himmetoglu *et al.*[31], who proposed the use of a "DFT+U+J" functional with U and J energy correction terms separately applied using calculated U and J parameters, together with a method for computing J. Linscott *et al.*[36], working within the minimum-tracking linear-response formalism, arrived at a factor-of-two smaller definition of Hund's J required for consistency with the definition of the Hubbard U. This latter definition has been shown to provide very accurate bandgaps for MnO[36] and $TiO_2$[37] when used with the DFT+U+J functional of Ref. [31]. In the present work, we combine the lessons of Refs. [31,36,37], calculating the Hund's J using the conventional Self-Consistent Field (SCF) finite-difference linear-response approach (instead of minimum-tracking) as available in standard Quantum ESPRESSO[38], but using the latter pre-factor for Hund's J as required by consistency considerations, as we will explain.

This functional will be referred to in this paper as the "DFT+U+J" following the nomenclature of Ref. [31], with the term emphasizing that the U and J energy corrections are both added seperately to each other as two distinct terms in the total energy. We emphasize that there is a long history of incorporating Hund's J parameter terms in the DFT+U family of functionals, indeed going back to the very origins of Hubbard-corrected density-functional theory. This wider class of functionals, including but certainly not limited to the functionals introduced in Ref. [11] and Ref. [35], are usually referred to by the names given to them by their originators, e.g. "LSDA+U", and we emphasize that by using the term "DFT+U+J" we refer exclusively to the functional given that name in Ref. [31]. There have also, of course, been many other extensions and applications of the basic concept, such as using different double-counting corrections[39], more modern functional forms[40,41], constrained RPA methods[42,43], and increasingly calculations including inter-site +V terms are becoming widespread.[44] These diverse techniques are all worthy of further investigation, but are outside the scope of the present study. Our findings are naturally not expected to be generalizable to any other functional including Hund's J apart from the two that we explicitly test, the well known Dudarev functional termed here DFT+(U-J), and DFT+U+J.

The simplified rotationally invariant DFT+U+J functional of Ref. [31] can be easily incorporated into any code that already is already capable of DFT+U, and indeed several codes including Quantum ESPRESSO's PWscf[45] and ONETEP[46] already support DFT+U+J. This functional results, in effect, from a rederivation of the Dudarev functional Eq. (2) in which it is argued that there is an additional energy term Eq. (3), proportional to Hund's J, that is consistent to retain given the starting assumptions and approximations. This term couples the subspace occupancy matrices of opposite spins, and survives when the (conventional fully-localized limit) double-counting correction is applied. It is furthermore argued in Ref. [31] that its double-counting approximation should be neglected, because it can't be represented in terms of the usual occupancy matrices and is unlikely to be well represented in the underlying functional to begin with. This removes a numerically problematic 'minority spin' term from the final DFT+U+J functional. We refer the reader to the original Ref. [31] for a comprehensive and very clear account of how the rotationally invariant DFT+U+J functional is derived, the extra physical mechanism of spin-flip exchange (which is beyond Hartree-Fock, and therefore arguably classifiable as correlation [32]) that it captures in addition to those effects in DFT+(U-J), and its expected effects in practical calculations (most notably, boosting local moments in spin-symmetry broken phases).

For non-spin-polarized systems, it is sufficient to use a DFT+U code to run DFT+U+J, as shown by Orhan *et al.*[37] and as we explain in section 2.1. Moreover, by taking this special case of the spin-polarized minimum-tracking linear-response framework proposed by Linscott *et al.*[36], it was found that the linear response U and J parameters can be computed simultaneously for non-spin-polarized systems. This method, termed here 'the γ-method', is described in the section 2.4 and is confirmed in the present study to work just as well within the well-known and long-established 'SCF' linear-response framework[13,14]. The γ-method is used extensively in the present work to calculate the Hund's coupling J as an effectively cost-free by-product of Hubbard U calculations.

While the DFT+U+J method is promising, it has not been evaluated on a large number of materials, as for example has recently been done for DFT+U with ortho-atomic orbitals by Kirchner-Hall *et al.*[27]. Furthermore, while there has been a vast number of studies involving DFT+U family methods, including Hund's J, and how they affect relaxed ionic geometries, there has also been little investigation into whether the DFT+U+J methodology (specifically referring to the functional of Ref. [31] with first-principles linear-response parameters) may introduce spurious geometric distortions into the lattice, as has been known to occur with other techniques employing U or J corrections, for example in Refs. [18-20,47]. This also partially motivates the present work. In this study, we have selected five representative $d^0$ or $d^{10}$ oxides as a challenging test set on which to benchmark and evaluate the DFT+U+J methodology against more common correction approaches. The materials chosen are $TiO_2$, $ZrO_2$, $HfO_2$, $Cu_2O$ and ZnO. These materials represent a range of different crystal structures, band-edge characters, and band-gap values, and have been previously studied extensively in experimental and computational literature. In order to ensure an accurate and fair evaluation of functionals, this transition-metal oxide test set was selected in advance of the study, and we report in detail separately on each material.

In this study, U and J parameters are calculated using the first-principles SCF linear-response method for each of the test set materials. The resulting values are then used within DFT+U+J simulations to evaluate the effect of these corrections on band-gaps, effective masses, and cell geometries. These results are compared against longer-established methods such as the Perdew-Burke-Ernzerhof (PBE) functional[48], DFT+U and DFT+(U-J) based on a PBE starting point, as well as against hybrid-functional results from the literature. We find that DFT+U+J yields band-gap accuracies similar to hybrid functionals such as HSE06, on average, without causing spurious distortion of crystallographic or band-structure parameters.

## 2. Methodology

### 2.1 DFT+U+J functional corrections

The Hubbard U correction is an additional energy contribution that is used to approximately correct for the many-body self-interaction error (SIE), or more generally delocalization error, that is harbored within specific pre-defined subspaces in a practical local or semi-local DFT calculation. The DFT+U energy correction term takes the form of

$$E_U[\hat{n}^{I\sigma}] = \sum_{I\sigma} \frac{U^I}{2} \mathrm{Tr}\left[\hat{n}^{I\sigma}(1 - \hat{n}^{I\sigma})\right], \quad (1)$$

where $\hat{n}^{I\sigma}$ represents the projected Kohn-Sham occupancy matrix for spin $\sigma$ and the subspace indexed $I$, and the U is an energy value that sets the magnitude of the correction. This reduces the degree of delocalisation by adding an energy penalty for non-integer occupancy matrix eigenvalues in the chosen subspace, which is spatially localized.

The U value can be selected by various methods, such as fitting to experiment, or reflecting past literature values. The linear response method, covered in the next section, allows for this value to be calculated from first-principles in situ for the material of interest, even self-consistently so[15], removing the reliance on empirical values.

Hund's J may be thought of as another correction parameter, on the same expansion order as the U and arguably a required counterpart to the U[36], which corrects for exchange effects that are ill-described by the approximate local or semi-local functional. The inclusion of J within DFT+U+J tends to promote high-spin states[31], however the effect remains relevant to the energy and potential of ultimately non-spin-polarized systems[37]. Indeed, in the Hubbard Hamiltonian context, the Hund's exchange contribution may be separated into terms that are quadratic separately in the total density and spin density[49].

The J parameter is commonly applied by combining U and J into a U-J parameter, as in the formalism of Dudarev *et al.*[34], which is a rotationally invariant simplification of previous formulations from Ref. [11] and Ref. [35]. In this method, which we will refer to as DFT+(U-J), the J value is simply subtracted from the U value and implemented as an "effective U" of (U-J), giving

$$E_{U-J}[\hat{n}^{I\sigma}] = \sum_{I\sigma} \frac{U^I - J^I}{2} \mathrm{Tr}\left[\hat{n}^{I\sigma}(1 - \hat{n}^{I\sigma})\right]. \quad (2)$$

However, there are also ways to implement the J correction separately, as an explicit term coupling unlike-spin densities. The DFT+U+J functional of Ref. [31] adds on a further term of the form (we neglect the 'minority term', as in that work and as is becoming customary)

$$E_J[\hat{n}^{\sigma}] = \sum_{I\sigma} \frac{J^I}{2} \mathrm{Tr}\left[\hat{n}^{I\sigma}\hat{n}^{I-\sigma}\right], \quad (3)$$

where the J is an energy value that scales the magnitude of the exchange correction applied to a subspace $I$. This superscript is dropped in subsequent equations for the sake of readability. This implementation was shown in Refs. [36] and [37] to provide accurate bandgaps with first-principles parameters for MnO and $TiO_2$, respectively.

The DFT+U+J functional evaluated in this work can be activated directly using some codes such as Quantum ESPRESSO and ONETEP, but as was shown in Ref [37], in the specific case of non-spin-polarized systems (when unperturbed) the DFT+U+J functional can be invoked in codes with no Hund's J implementation at all. This involves replacing the applied U parameter with an effective U of $U_{full} = U - 2J$, and simultaneously applying a subspace potential shift of $\alpha = J/2$. As shown in that paper, the result is mathematically identical (in the energy and its derivatives, e.g., potential, forces, etc.) to applying separate U and J corrections with the original values defined in DFT+U+J (the combination of Eq. (2) plus Eq. (3).

### 2.2 Fundamental considerations: Accessibility of band gaps and the paired nature of Hubbard U and Hund's J corrections

As this work focuses principally on the predictive capacity of DFT+U and DFT+U+J for the fundamental band gap, in comparison to other methods such as hybrid functionals, it is worth considering carefully whether this quantity is accessible in principle. For most of the history of practical Kohn-Sham DFT calculations, although the discussion of the Kohn-Sham gap was commonplace, it was supposed that any agreement with experiment was, at best, fortuitous. This is because, while DFT is a ground-state theory, the fundamental gap is a charged excited-state property that formally requires the removal and addition of electrons.

In recent years, however, it has become apparent that the fundamental band gap is a valid quantity to associate with the single-particle eigen-spectrum from a DFT calculation after all, but only under specific conditions. Namely, when considering a solid with a well-converged k-point sampling, and when using not Kohn-Sham DFT but *generalized* Kohn-Sham (GKS) DFT, then it can be shown that an explicit addition-and-removal calculation will give the same fundamental band-gap as that extracted from the gap in the GKS eigen-spectrum. A GKS calculation requires the orbitals to provide the domain (degrees of freedom) for the calculation, not the density. Rather than being an added complication, in practice this has been the most commonplace mode in which DFT is used, since the very beginning. For this distinction to have discernible effect, the potential needs to be non-multiplicative, i.e., it should be a non-local potential such as that provided by DFT+U, DFT+U+J, or indeed hybrid functionals.

A GKS calculation is not enough to open the eigen-spectrum gap to a reasonable value, which in density-only DFT requires the post hoc addition of the derivative discontinuity to the eigen-spectrum gap. Several families of orbital-dependent functionals, however, including hybrids and DFT+U, can incorporate a reasonably good approximation to the necessary derivative discontinuity directly with their reference system (which, unlike that of Kohn-Sham DFT, is partially interacting), so that the effect of such interactions manifests directly in the GKS eigen-spectrum gap. As a DFT+U, DFT+(U-J), or DFT+U+J calculation (on an extended system) meets the necessary conditions, the fundamental gap from such a calculation is both comparable in principle and in practice to the fundamental band-gap, noting of course that there would be several additional effects required for perfect comparability, such as zero-point and finite-temperate phonon corrections, which are not typically included. As the abstract considerations mentioned here are beyond the scope of the present study, for further details on the domain of physical validity of GKS band gaps, we refer the reader to the more comprehensive Refs [1,50-56].

The abstract justifiability of a given formalism for calculating the band gap does not, of course, imply that it will do so accurately, numerically speaking. This brings us to the question of why it may be advantageous, or even in principle necessary, on physical grounds, to incorporate Hund's J corrections on the same footing as Hubbard U corrections. On practical grounds, computing and incorporating the J is cost-free when using the 'gamma' method discussed in Section 2.4, but that is a somewhat lesser consideration. Within the spin-polarized linear-response formalism as detailed in Ref. [36], it becomes clear that the J is a parameter on perfectly the same footing as the Hubbard U, in terms of powers of the spin-density and pairing of orbital indices. Like the Hubbard U, the Hund's J within the present context measures, for pre-defined subspaces, a supposed pathology of the approximate functional. This pathology is an energy-magnetization curvature normally referred to as 'static correlation error'[52] notwithstanding that exchange mechanisms may also give rise to magnetization-squared behavior [33]. Together, the Hubbard U and Hund's J measure different aspects[57] of the deviation from the well-known flat-plane condition of exact DFT[54]. Specifically, U and J are computed while supposing that the flat-plane condition holds approximately for the subspace undergoing analysis, and by ensuring that curvature contributions that are not directly related to interactions are subtracted off from their values (as interaction is to be corrected, so only interaction is measured). Correspondingly, it is shown in Ref [31] that the interaction terms that U and J pre-multiply also appear on the same order, in terms of powers of the spin-density and pairing of orbital indices. The Hubbard U and Hund's J can be thought of as 'two sides of the same coin' of spin-indexed linear-response theory or, rather more precisely, as measuring the curvatures across two different slices of the surface formed by the interaction energy as a function of both the subspace occupancy and subspace spin-magnetization.

What our results will show is that, while the spin magnetization naturally vanishes in closed-shell oxides, the energy-magnetization *curvature* measured by J can remain large, even for O 2p orbitals. This is a feature of the approximate functional under scrutiny, and it is an indicator of its ill-description of multi-reference effects. In summary, there is a non-zero Hund's J type error to be measured for this functional for these closed-shell oxides. Where it comes to using this J, we rely on the landmark work of Ref. [31], where it was shown that a spin-off-diagonal term proportional to energy-magnetization curvature survives upon re-derivation of the long-established simplified rotationally invariant DFT+U functional. This leads to the first-principles 'DFT+U+J' functional that we use, which ultimately corrects for spin-flip pair hopping. This correlation effect [32], which is necessarily approximated in terms of 1-body occupancy matrix products, is different and additional to same-spin pair-hopping exchange that is already contained and approximated in the same way within the DFT+(U-J) functional.

Let us look next at how the addition of the Hund's J modifies the derivative discontinuity and hence the GKS prediction of the fundamental gap. For a Mott-Hubbard material with perfect projection of the same subspace type at both valence and conduction band edges, such as that provided by 3d-like Wannier functions, and neglecting self-consistent effects, the derivative discontinuity and band-gap opening will be exactly the corresponding Hubbard U. Using the DFT+(U-J) functional this is of course reduced to U-J (reduced if we assume that the J is positive, as it ordinarily is). As shown in Ref. [37], in DFT+U+J the gap opening would fall further to U-2J, in the absence of magnetization. Taking a more realistic case of vanishing magnetization, but looking instead at a charge-transfer insulator like $TiO_2$, we may suppose perfect O 2p projection at the valence band and perfect Ti 3d projection at the conduction band. Again ignoring self-consistent effects and other types of bands that may be present, the band gap opening will go like the average of the O 2p and Ti 3d U values in DFT+U, and the average of their U-J values in the case of DFT+(U-J). The situation is more complex in DFT+U+J due to the occupancy independent subspace potential shift of $\alpha = J/2$, previously mentioned. In this case, the Hund's J value of the valence band becomes more dominant, and the band-gap will be the average of the U-2J values of the two band-edge subspaces, together with a further correction (typically a further reduction) of size $(J_{3d} - J_{2p})/2$.

The typically opposing effects of the Hubbard U and Hund's J on the band gap can be rationalized further by considering that positive U indicates convexity in the interaction energy landscape, while positive J indicates concavity in the same landscape. These effects, namely delocalization error and static correlation error, respectively, are known to cancel out to a considerable extent in common semi-local functionals for many systems. Therefore, while addressing one without the other might give a better result for a particular system, the cancellation of error in the energy can be expected to be less reliable more generally. A derivative discontinuity in the energy is needed to maintain good boundary conditions while removing spurious curvatures, and indeed DFT+U type functionals provide this. As this provides the GKS band-gap and hence fundamental gap prediction, the same argument again follows. Specifically, we can expect that by addressing one error only, by applying a Hubbard U in DFT+U, or by also but only partially addressing the other error, as in DFT+(U-J), we can expect to leave a systemic error in the band-gap that is only resolved when both are treated on the same footing, as in DFT+U+J. We emphasise finally in this section, however, that the DFT+U+J functional tested here is far from the only one in use, and it already does not represent the last word in the development Hubbard corrective functionals using U and J.

### 2.3 Linear response calculations of U and J

The total energy in subspace-perturbed DFT is given by

$$E = E_{\mathrm{DFT}} + \alpha N + \beta M, \tag{4}$$

where N is given by N = $n^{\uparrow} + n^{\downarrow}$, where M is given by M = $n^{\uparrow} - n^{\downarrow}$, and where the $n^{\sigma}$ are the traces of the perturbed subspace occupancy matrix of each spin. $\alpha$ and $\beta$ are the strengths of the subspace-uniform potentials that allow the occupancy N and magnetization M, respectively to be controlled at minimum energy cost, as guaranteed by the properties of the fully-relaxed constrained DFT energy landscape[58]. Here, $\alpha$ is the strength of a perturbation that is the same for both spins. In the case of controlled magnetization M, the perturbation is repulsive with strength $\beta$ for spin-up and attractive with strength $\beta$ for spin down. Therefore, $\beta$ is *half* of the difference in perturbation between the two spin channels, and we will recall this factor of one-half again presently.

To calculate U, a number of different $\alpha$ potentials are applied, and the linear response of the occupation numbers gives the values for the bare (i.e., unscreened, or one-shot) $\chi_0$ and the relaxed $\chi$ subspace-projected linear response:

$$\chi_0 = \frac{dN_0}{d\alpha}, \chi = \frac{dN}{d\alpha} \tag{5}$$

Here, $N_0$ is the total occupancy trace for the bare case. Fig. 1a) shows example calculations of $\chi_0$ and $\chi$, with the slopes being calculated from least-squares linear regression. The U value can then be calculated [13,14] for on-site only DFT+U type corrections using the scalar equation:

$$U = \chi_0^{-1} - \chi^{-1} \tag{6}$$

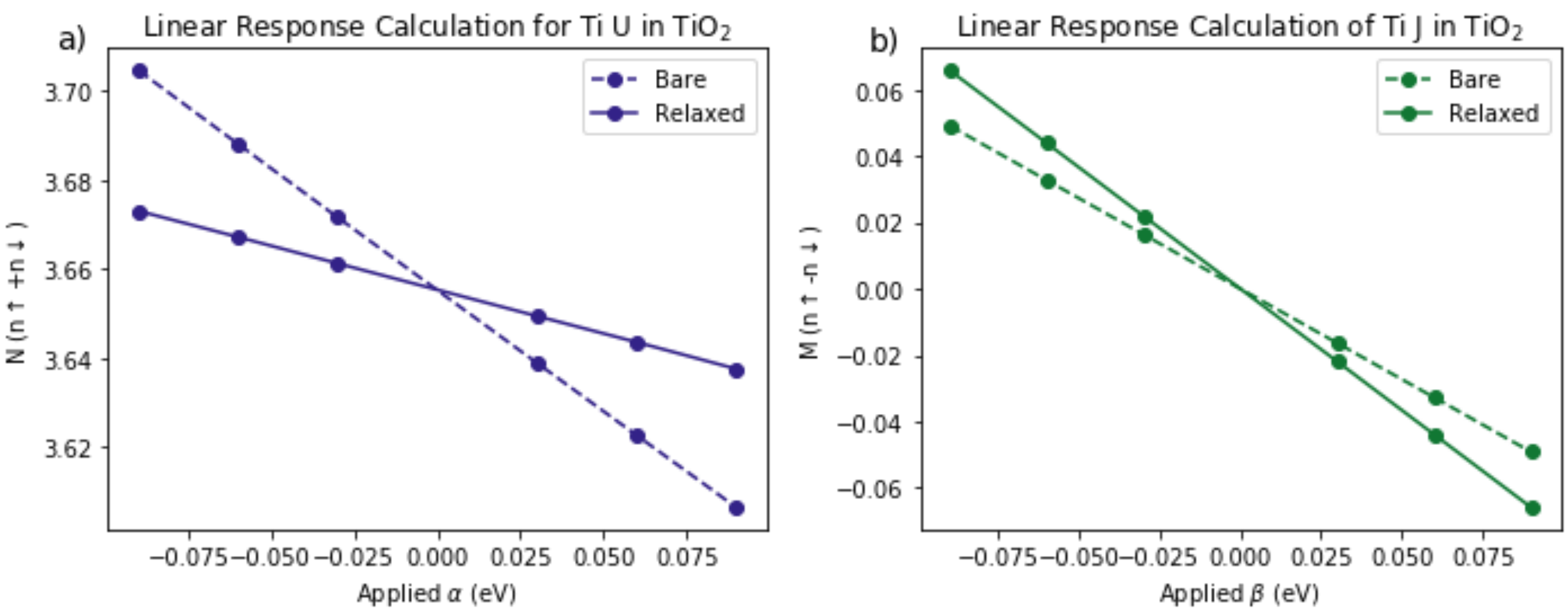

**Fig. 1. Example linear response calculations of a) the Hubbard U and b) Hund's exchange J for a pseudo-atomic Ti 3d orbital subspace in rutile $TiO_2$.**

In order to see how this approach can be adapted to calculate the Hund's exchange coupling J, we base our definition on the ansatz for J for minimum-tracking linear response as defined in Eq. (22) of Linscott *et al.*[36], specifically

$$J = -\frac{1}{2} \frac{d(v_{\mathrm{Hxc}}^{\uparrow} - v_{\mathrm{Hxc}}^{\downarrow})}{d(\mathrm{n}^{\uparrow} - \mathrm{n}^{\downarrow})} \tag{7}$$

This formula measures the rate of change, with respect to subspace magnetization M, of subspace-averaged part of the interacting potential (Hxc denotes Hartree, exchange, and correlation) that applies to the magnetization density. This is akin, but technically different, to minus the second derivative of the interaction energy with respect to M, which would represent the global rather than subspace-specific analogue. Spurious energy-magnetization curvature is well associated with static correlation error in approximate density-functional theory, and we can interpret Hund's J as a measure of this, at least within the approximate subspace-bath decoupling and screening approximation of that underpins DFT+U type correction and linear response.

It is important to note at this point that this formula for J differs by a factor of one-half with respect to the similar formula proposed in Himmetoglu *et al.*[31], which was derived by analyzing the double-counting correction of the DFT+U+J functional introduced in that work, which substantially inspires the present one. The factor of one-half brings consistency with the Hubbard U as shown in Linscott *et al.*[36] (see Eqs. 20 and 24), by ensuring that the interacting analogue of the perturbation strength (the magnetization Lagrange multiplier $\beta$) is the potential that is being measured, and not twice that. Eq. (7) has recently used to calculate very accurate $TiO_2$ band-gaps from first-principles in Ref. [37]. The global minus sign is a matter of long-standing convention, e.g., from the Ising and Heisenberg Hamiltonians. A further justification of the factor of one-half in Eq. (7) is offered in Appendix I, which relies on the analysis of a toy system rather than the relatively involved prior analysis of Ref. [36].

While this formula could, in principle, be directly used within self-consistent field DFT codes, in this work we explore instead its analogue within the long-standing approach used in those codes for the non-interacting, bare response, namely that of evaluating perturbed occupancies before the interacting part of the potential begins to be updated. To see this, we first can rewrite

$$J = -\frac{1}{2}\frac{d((v_{\mathrm{KS}}^{\uparrow} - v_{\mathrm{ext}}^{\uparrow}) - (v_{\mathrm{KS}}^{\downarrow} - v_{\mathrm{ext}}^{\downarrow}))}{d\mathrm{M}} = -\frac{1}{2}\frac{d(v_{\mathrm{KS}}^{\uparrow} - v_{\mathrm{KS}}^{\downarrow})}{d\mathrm{M}} + \frac{1}{2}\frac{d(v_{\mathrm{ext}}^{\uparrow} - v_{\mathrm{ext}}^{\downarrow})}{d\mathrm{M}} = -\frac{d\beta_{\mathrm{KS}}}{d\mathrm{M}} + \frac{d\beta}{d\mathrm{M}} \approx -\frac{d\beta}{dM_0} + \frac{d\beta}{d\mathrm{M}}. \tag{8}$$

Here, $\beta_{KS}$ is the average of the spin-affecting Kohn-Sham potential over the subspace following the framework of Refs. [37,58], and the approximate equality signifies the relationship between minimum-tracking and SCF linear-response. Note that Refs. [36,37] use the minimum-tracking method, while this study uses only conventional SCF linear response in the spirit of Ref. [13], and specifically we use the scalar (not matrix) equations (21-22). In practice, notwithstanding, within self-consistent field codes such as Quantum Espresso, Abinit, CASTEP, VASP, and others, calculating the value for J from linear response involves calculating the bare and relaxed response of M to an applied $\beta$, as shown in Fig. 1b:

$$\chi_{M0} = \frac{dM_0}{d\beta}, \chi_M = \frac{dM}{d\beta}, \tag{9}$$

and J can ultimately be calculated, within the SCF linear-response formalism, as:

$$J = -\chi_{0M}^{-1} + \chi_{M}^{-1}. \quad (10)$$

We note that relaxation tends to enhance the magnetization response, in contrast to the occupancy response which is always reduced by screening (when perturbing from a stable state[58]). A positive value for the computed J indicates an erroneous effective magnetization-magnetization interaction within the subspace with a sign corresponding to underestimated Hund's exchange coupling in the underlying functional.

**2.4 The γ-method for simultaneous U and J calculation**

Orhan and O'Regan[37] have demonstrated that within the minimum-tracking linear-response formalism and for non-spin-polarized (when the potential is unperturbed) systems, a simple procedure can be used to calculate both U and J simultaneously, using half the number of finite-difference calculations with respect to the usual method. The same procedure should work for the methods used in this work, as we now investigate.

In calculations with both $\alpha$ and $\beta$ parameter, the $\alpha$ parameter is applied equally to each spin, while the $\beta$ parameter applies an opposite potential to each spin channel. In terms of the average values of the Kohn-Sham potential within the perturbed subspace[58], we have

$$V^{\uparrow} = V_{\mathrm{DFT}} + \alpha + \beta, \quad (11)$$

$$V^{\downarrow} = V_{\mathrm{DFT}} + \alpha - \beta. \quad (12)$$

If we set the value of $\alpha$ to be equal to $\beta$, then the result is that a potential of $2\alpha$ is applied to the spin-up channel alone, while the spin-down channel is unchanged. Setting $\gamma=2\alpha$ for notational convenience, we can determine the spin-indexed response matrix[36] components from the trace of each individual spin channel occupancy matrix:

$$\chi^{\uparrow\uparrow} = \frac{d\mathrm{Tr}[n^{\uparrow}]}{d\gamma} \quad (13)$$

$$\chi^{\downarrow\uparrow} = \frac{d\mathrm{Tr}[n^{\downarrow}]}{d\gamma} \quad (14)$$

For a non-spin-polarized system, the remaining components of $\chi$ are set by time-reversal symmetry:

$$\chi^{\uparrow\uparrow} = \chi^{\downarrow\downarrow} \quad (15)$$

$$\chi^{\downarrow\uparrow} = \chi^{\uparrow\downarrow} \quad (16)$$

These results together define the 2x2 response matrix $\chi$:

$$\chi = \begin{pmatrix} \chi^{\uparrow\uparrow} & \chi^{\uparrow\downarrow} \\ \chi^{\downarrow\uparrow} & \chi^{\downarrow\downarrow} \end{pmatrix} \quad (17)$$

This same procedure is separately used to determine the bare response $\chi_0$. The matrix difference between the inverted $\chi$ and $\chi_0$ matrices then yields the 2x2 interaction matrix $f$:

$$f = \chi_0^{-1} - \chi^{-1} \quad (18)$$

For ultimately non-spin-polarized systems (systems that are non spin-polarised in the absence of perturbations), where Eqs. (15,16) hold, the U and J values can then be derived from the elements of the resulting interaction matrix [36,37] as

$$U = \frac{f^{\uparrow\downarrow} + f^{\uparrow\uparrow}}{2} \quad (19)$$

$$J = \frac{f^{\uparrow\downarrow} - f^{\uparrow\uparrow}}{2} \quad (20)$$

Simplified into single scalar equations, and using the assumptions of Eqs. (15,16), the resulting equations become, in terms of the scalar quantities defined in Eqs. (13,14):

$$2U = \left(\chi_0^{\downarrow\uparrow} + \chi_0^{\uparrow\uparrow}\right)^{-1} - \left(\chi^{\downarrow\uparrow} + \chi^{\uparrow\uparrow}\right)^{-1} \quad (21)$$

And:

$$2J = \left(\chi_0^{\downarrow\uparrow} - \chi_0^{\uparrow\uparrow}\right)^{-1} - \left(\chi^{\downarrow\uparrow} - \chi^{\uparrow\uparrow}\right)^{-1} \quad (22)$$

An example of the calculations used to determine U and J with this method is shown in Fig. 2. Note that the bare spin down response is unaffected by γ, as in this method only the spin-up Kohn-Sham potential is modified with respect to its ground-state profile, with the spin down potential contains no perturbation and thus the spin-down density exhibits no non-interacting response.

Linscott *et al.*[36] have shown that within the minimum-tracking formalism, the calculated U and J from the scaled 2x2 methodology should match with $\alpha$ and $\beta$ method linear response U and J calculations. Following this, Orhan *et al.*[37] utilized the efficiency brought by time-reversal symmetry to calculate minimum-tracking U and J values for both rutile and anatase $TiO_2$ with an LDA starting point. The exploitation of time-reversal symmetry for ultimately non-spin-polarized systems could, in principle, allow for the calculation of both U and J with as few as two γ-point simulations per subspace, although more points are used in this study to ensure that the resulting response is linear.

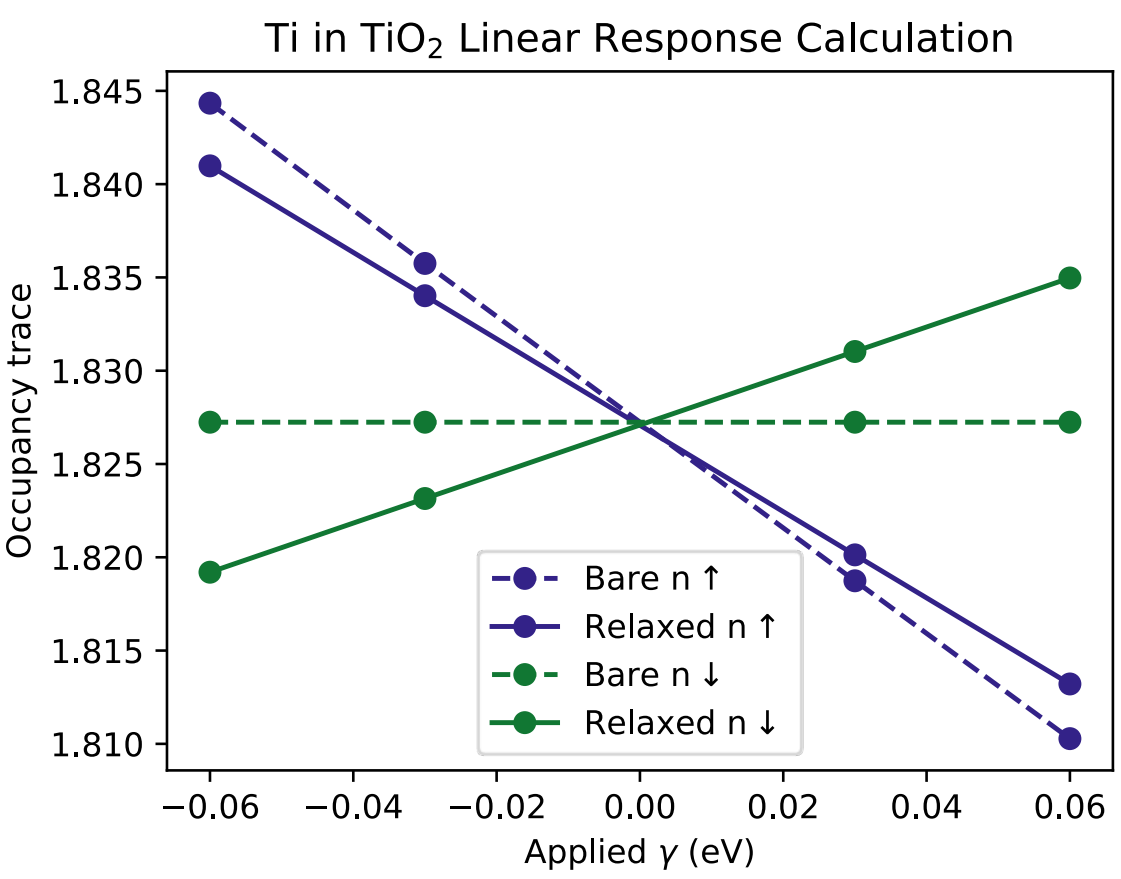

**Fig. 2. Example linear response calculation of U and J for a pseudo-atomic Ti 3d orbital subspace in rutile $TiO_2$, using the γ-method and the standard Quantum Espresso package.**

Before leaving the theoretical methods behind, we emphasize that the γ-method is only applicable to non-spin-polarized systems. The same is true, even if $\alpha$ and $\beta$ are separately used, for Eqs. (19-22). For spin-polarized systems, the calculation of J requires the use of Eq. (10) directly, and ordinarily perturbing M using $\beta$.

### 2.5 Corrective functionals incorporating U and J

In this study, four different methods for incorporating the calculated U and J are considered and compared. The first method, referred to as DFT+U (metal only), takes the bare PBE simulation and adds the U correction to metal-atom centered nd subspaces alone, which is the most common application of U to TMOs. The quantum number n here is one less than the period in which the metal element resides. The correction is applied using Eq. (1). The second method, referred to as DFT+U (all atoms), applies the calculated U corrections to both metal and oxygen 2p subspaces, a practice that has been used in many recent studies[21-27].

The J parameter can be applied in different ways. The third method investigated involves subtracting the J value for each atom from the U value and applying this U-J value as the new U correction on all atoms, referred to as DFT+(U-J). The final method is that outlined by Linscott *et al.*[36], where the DFT+U+J functional of Eq. (7) and Ref. [31] is used to apply U and J corrections separately to metal and oxygen atoms, referred to here simply as DFT+U+J.

For the sake of completeness, bandgaps were also calculated for the method of applying both U and J to the metal atoms alone and not the oxygen. However, in all materials except ZnO, the bandgap accuracy did not improve and its value was very close to that in the case of U on the metal atoms alone. As a result, these results such have been omitted from the figure and further analysis but can be found in Appendix II.

Each of these methods, alongside uncorrected PBE, are evaluated for their effect on the fundamental (generalized Kohn-Sham) band-gap, relaxed crystallographic parameters, and electronic band-structure of the test set materials, in order to determine whether they can accurately predict the bandgap of each material without causing distortions in other material parameters. It should be noted that these are strictly evaluations of these particular methodologies (corrective functionals and accompanying method for calculating U and J), and our results are not necessarily transferable to other techniques, however related.

### 2.6 Computational details

The calculations in this study were conducted using the Quantum ESPRESSO package[38], utilizing the PBE exchange-correlation functional. For each element in every material, the charge neutral PSlibrary 1.0.0[59] ultrasoft pseudopotentials were used. The unit cell for each material was converged with respect to kinetic energy cut-off, charge density cut-off, and k-points, until the variable-cell relaxed energy difference was less than 1 meV per atom. Table 1 shows the converged unit cell parameters for each of the five materials. The ratio of charge density energy cutoff to wavefunction energy cutoff was converged even at the minimum value of 4:1, likely due to the much higher-than-typical wavefunction cutoff value used out of an abundance of caution, i.e., to ensure that projector orbitals where fully sampled. In practical studies beyond the present careful benchmarking, it may be that computational efficiency can be optimized by using higher grid ratios but lower wavefunction cutoffs than those shown in Table 1. The energy and force convergence thresholds for relaxations were $6 \times 10^{-5}$ Ry and $10^{-4}$ Ry/Bohr respectively, a Fermi-Dirac smearing of 0.01 Ry was applied, and the Brillouin zone was sampled using a Γ-centered Monkhorst-Pack grid[60]. The Hubbard projectors were defined in the default way within PWscf, that is as neutral-configuration non-orthonormalized pseudo-atomic orbitals, except where for one material ($Cu_2O$) the effects of orthonormalization was tested, as described in Appendix III.

**Table 1: Converged simulation parameters for the unit cell of each of the five materials considered, meeting a 1 meV per atom convergence criterion.**

| Material | Wavefunction Energy cut off (Ry) | Charge density energy cut-off (Ry) | k-point grid |
|---|---|---|---|
| $TiO_2$ | 120 | 480 | 3x3x5 |
| $ZrO_2$ | 120 | 480 | 3x3x3 |
| $HfO_2$ | 130 | 520 | 3x3x3 |
| $Cu_2O$ | 60 | 240 | 5x5x5 |
| ZnO | 120 | 480 | 6x6x6 |

Before proceeding with calculations based on the γ-method, we first numerically verified its equivalence (for non-spin-polarized systems only) to the approach of separately using $\alpha$ for U and $\beta$ for J, using the non-supercell-converged unit cells of each of our test materials.

The first practical step in each case was to save the wavefunctions of a single-point SCF calculation using an energy convergence threshold of $10^{-6}$ Ry. This was then restarted with the same initial wavefunction but with applied $\alpha$ (for U calculations), $\beta$ (for J calculations) potentials along with tighter convergence thresholds of $10^{-11}$ Rydberg for the initial diagonalization and $10^{-9}$ Ry total-energy convergence. The trace of the bare and relaxed occupancy matrices was extracted for several different applied potentials. Least-squares linear regression was used to determine $\frac{dN}{d\alpha}$ and $\frac{dM}{d\beta}$ and thus extract the appropriate $\chi$ and $\chi_0$ values to be used in Eq (6) (to calculate U) and Eq. (10) (to calculate J). The procedure for the $\gamma$-method is similar, with a range of applied $\gamma$ values being used to calculate $\frac{dn}{d\gamma}$ for the spin-up and spin-down channels, and thus calculate the required $\chi$ and $\chi_0$ values that can be used to calculate U and J using Eqs. (21,22).

### 2.7 Effective mass calculations

The band-structure of each material between selected high-symmetry points was calculated in order to determine the location of the band-gap, and whether it is direct or indirect. In order to evaluate the effect of each method on the band-structure of the material, the scalar path-dependent effective mass was calculated from

$$m^* = \pm \hbar^2 \left( \frac{d^2 E}{dk^2} \right)^{-1}. \quad (23)$$

The sign is positive when evaluating electron effective mass and negative when calculating hole effective mass. This was calculated from a parabolic fit at an appropriate energy range above or belove the band edge for electrons and holes, respectively. The ranges in which a reliable parabolic fit was extracted were 0.05 eV for $TiO_2$, $HfO_2$ and $Cu_2O$; 0.01 eV for $ZrO_2$, and 0.1 eV for ZnO.

## 3. Results

### 3.1 Verification and evaluation of the γ-method

**For the purposes of validating the γ-method, which halves the cost of calculating the parameter pair for non-spin-polarized systems, U and J values were calculated for the unit cell of each material. We compared the results given by the $\alpha$ and $\beta$ method of Section 2.2 and the γ-method of Section 2.3.**

Table 2 shows a comparison of the resulting values for rutile $TiO_2$, which agree to within 0.5% of the U and J values calculated in the more obvious way, that is with $\alpha$ for U and $\beta$ for J.

**Table 2: Demonstration that the U and J values calculated for $TiO_2$ with the $\alpha$ and $\beta$ method of linear response and with the γ-method involving simultaneous U and J calculation are identical.**

| Parameter | $\alpha$ and $\beta$ method (eV) | γ-method (eV) | Difference (eV) | Difference (%) |
|---|---|---|---|---|
| Ti U | 3.238 | 3.228 | -1.087E-02 | -0.336 |
| Ti J | 0.465 | 0.465 | 3.247E-04 | 0.070 |
| O U | 12.070 | 12.035 | -3.519E-02 | -0.292 |
| O J | 1.826 | 1.835 | 8.823E-03 | 0.483 |

Similar tables for the other four materials are collected in Appendix IV. For all atomic elements of all materials, there is less than 1% difference between the U and J calculated using the methods, indicating that that the γ-method is equivalent to the $\alpha$ and $\beta$ method for ultimately non-spin-polarized systems. In Ref. [36], this was confirmed to be the case also within the minimum-tracking linear response definitions of U and J. By using the γ-method, when spin polarization is not anticipated, the calculation of Hund's coupling J can be performed as an essentially cost-free by-product of calculating the Hubbard U, using Eqs. (21,22). Thus, it is encouraging to confirm its validity here for the SCF (standard) linear-response calculations now very routinely performed using plane-wave DFT codes.

### 3.2 Rutile $TiO_2$ results

The linear response U and J values for rutile $TiO_2$ were calculated using different supercell sizes, as shown in Table 3. These parameters were performed for the Ti 3d and O 2p orbitals separately, as motivated by previous studies [25-27,37,61,62]. At acceptable convergence of the derived quantities that these parameters will be used to calculate, the supercell calculations ultimately agree to within 0.04 eV for all parameters. The calculated U values for Ti and O are within the range of previous linear response studies[25,27,37]. It should be noted that different population analysis schemes for defining the DFT+U subspaces can yield significantly different calculated U parameters and subsequent results[27,63]. For example, the non-orthonormalized Hubbard projectors used in this study will produce different U values than that of orthogonalized projectors.[27] The U and J values for the largest supercell size of 3x3x5 (270 atoms) were used for the materials property calculations that follow.

It is worth noting that rather satisfactory convergence would already be recovered using smaller cells, noting the evident erratic nature of the convergence profile but also the insensitivity of the predicted properties with respect to small changes in the parameters. This rapid convergence reflects, perhaps, our use of scalar rather than matrix response inversion, following Refs [36] and [37] and using Eq. (21 and 22), which is predicated on the assumption that all atoms but the perturbed one should participate in screening when no +V parameters are to be calculated. The more rapid supercell convergence with little effect on the converged U value, due to scalar inversion, has been identified by other authors, e.g., in Ref [64], and is an avenue for further study outside the scope of the present work. It is further interesting to note that the convergence rate is not the same for all orbital types, or between U and J. The spatial extent of the projector orbitals, as depicted in Fig. 8, play a complex role, as does as the anisotropic non-local screening environment. We can discern no overall trend in convergence rate across the materials studied. In methods such as constrained RPA, where J is defined to capture only explicit exchange integral matrix elements, the J is well known to converge rapidly and to be relatively impervious to the screening environment. Here, however, we find that the linear-response density-functional theory J (interpreted as a localized measure of spurious magnetization-magnetization interaction energy, which is not limited to exchange interactions alone and indeed usually termed static correlation error) from Eqs. (7, 8 or 22), converges similarly to the Hubbard U, reflecting their inter-related defining formulae.

**Table 3: Convergence of calculated linear response U and J values for Ti 3d and O 2p subspaces in rutile $TiO_2$ with different supercell sizes and k-point grids.**

| Supercell size | k-points | Ti U (eV) | Ti J (eV) | O U (eV) | O J (eV) |
|---|---|---|---|---|---|
| unit cell | 5x5x8 | 3.228 | 0.465 | 12.035 | 1.835 |
| 2x2x3 | Γ-point | 3.238 | 0.383 | 11.214 | 1.679 |
| 2x2x3 | 2x2x2 | 3.240 | 0.385 | 11.237 | 1.682 |
| 3x3x5 | Γ-point | 3.225 | 0.384 | 11.199 | 1.688 |

Experimentally, the fundamental band-gap of rutile $TiO_2$ has been measured to be 3.03 eV by very-high-resolution absorption[65] and time-integrated photoluminescence[66]. Fig. 3a) shows a comparison of rutile $TiO_2$ bandgaps derived from experiment, from hybrid and GW techniques, and from the methods examined in this study. The average HSE06 bandgap value[65,66] is the closest of the hybrid methods, with a slight bandgap overestimation of 0.24 eV, while PBE0 values[67,68] found in the literature overestimate the gap by an average of 0.92 eV, and even $G_0W_0$ methods[69,70] overestimate by 0.37 eV. Our PBE value with no corrections applied greatly underestimates the experimental gap by 1.19 eV, and applying DFT+U to the Ti 3d subspace alone does not significantly improve this, in agreement with previous studies[27,37,62]. When the U is applied to both Ti and O, however, the bandgap is increased and becomes 0.80 eV larger than the experimental one.

Incorporating Hund's exchange coupling J as well as U brings the calculated bandgap back down. In the case of the DFT+(U-J) method, there is still an overestimation of 0.39 eV, while the DFT+U+J functional is the most accurate of those tested, giving an underestimation of 0.24 eV. This echoes the recent findings for rutile in Ref. [37], albeit that a significantly better agreement with experiment was found there with an LDA rather than a PBE starting point, and with charge-neutral LDA pseudo-atomic orbitals defining the subspaces. The zero-point phonon correction is expected to be negligible in rutile[71], with respect to the band-gap inaccuracies in question here. The DFT+U+J bandgap error here is lower than both PBE0 and $G_0W_0$ methods and is equal to the HSE06 average error, but with a significantly lower computational cost and indeed minor extra cost over PBE after the U and J parameters are calculated.

To give an indication of the effect of convergence on U and J magnitudes, the bandgap was also calculated for the unconverged unit cell parameters from Table 3. This had slightly higher U and J values (most especially the 0.8 eV higher O U), yielding a bandgap of 2.834 eV, which is an increase of only 0.042 eV. This is an indicator that the level of convergence in the U parameter does not need to be especially high to give stable values for the bandgap.

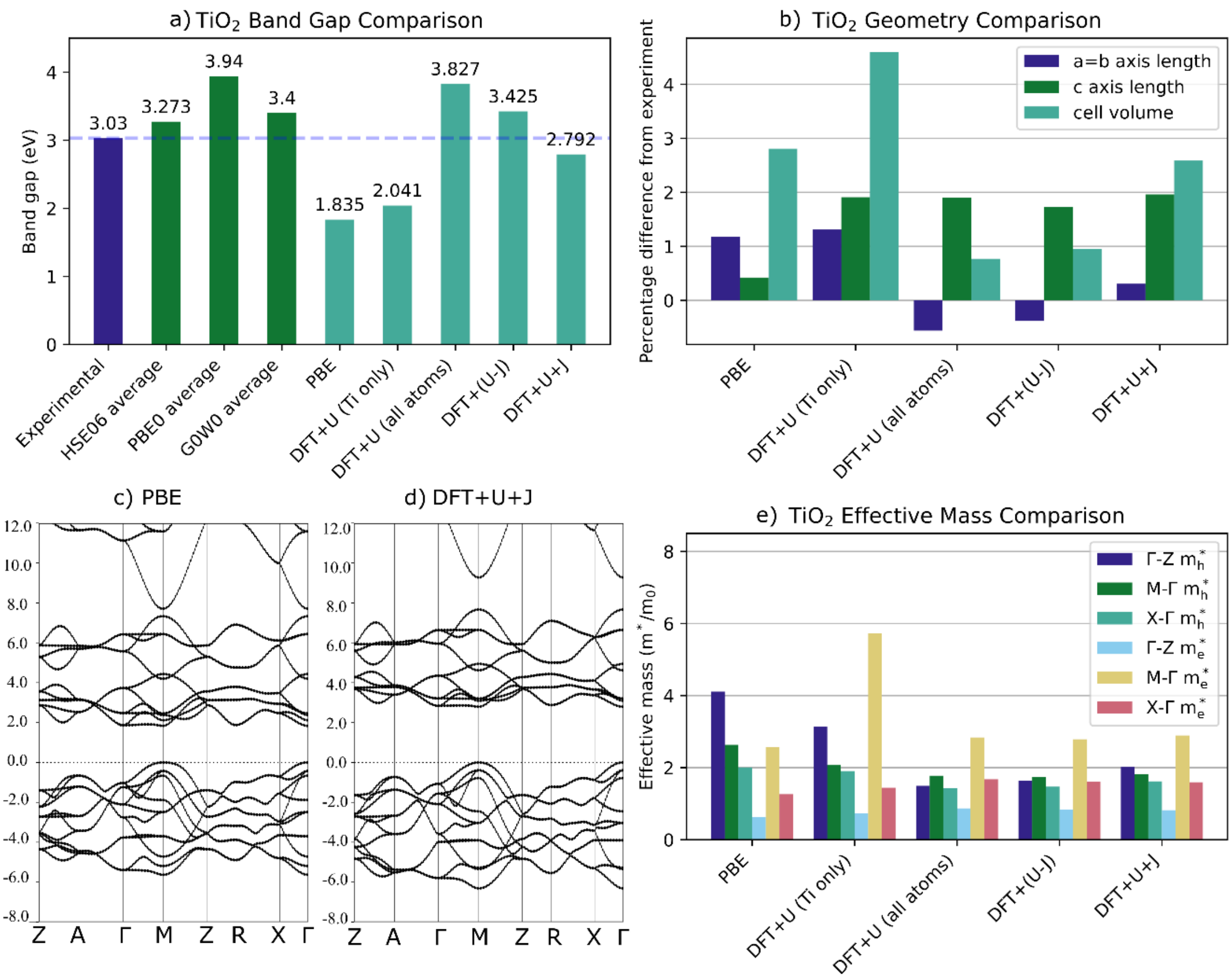

**Fig. 3. Summary of effect of different U and J incorporating corrective functionals on rutile $TiO_2$ material properties. a) Comparison of the experimental rutile bandgap[65,66] with an average of literature bandgap values for HSE06[70,72,73], PBE0[67,68], and $G_0W_0$[69,70], and with the five PBE-derived functionals utilizing U and J parameters. b) Percentage deviation of crystallographic parameters (axis lengths and volume) from experimental values for each correction method. c) The band-structure of rutile for the PBE functional with no corrections applied. d) The band-structure of rutile using the DFT+U+J functional. e) Calculated effective mass ratios in selected directions for the five functionals.**

Fig. 3b. shows the effect of the methodologies on the axis lengths and volume of the cell, as a percentage of the experimental value. Applying the +U on the Ti 3d orbitals results in a stretching of the c-axis, increasing the volumetric error by around 2%. Applying DFT+U on both the Ti 3d and O 2p orbitals maintains the c-axis stretching, but shrinks the a and b axes to below the experimental value, resulting in a volumetric error that is smaller than that of uncorrected PBE. Applying U and J, both to Ti 3d and O 2p subspaces results in an a=b axis length very close to experiment, but slightly increases the volumetric error. Overall, apart from the U on Ti 3d only case, no methodology results in a significantly distorted geometry with respect to either experimental or PBE values. This is reminiscent of the findings in Linscott *et. al.*[36], where it was shown that the O 2p counterpart correction cancels the tendency for metal 3d correction to over-elongate bonds in hydrated metal complexes.

There is similarly little band-structure distortion arising from the DFT+U+J methodology. Fig. 3c. and Fig. 3d. depict the band-structures for bare PBE and for the full DFT+U+J methodology. There does not appear to be a large structural difference in the band-structure near the gap, apart from the greater bandgap, with both methods predicting a direct bandgap with the conduction-band minimum (CBM) at the Γ point, only slightly lower in energy than the CBM at the M point. This difference is 32 meV for PBE and 45 meV for DFT+U+J. The bandwidth of the valence band is also increased from 5.6 eV for PBE to 6.3eV for DFT+U+J.

Fig. 3e. shows how the calculated effective mass of electrons and holes were affected by the corrective functionals. The application of DFT+U tends to decrease the effective mass of holes, with the largest effect occurring in the Γ-Z direction. The effective mass of electrons is increased when U is applied, and this is most prominently seen along the M-Γ direction for the DFT+U (Ti only) method, although this may be an artifact of the very flat band-structure along this direction. The U corrections also appear to have the effect of making the band-structure of the holes more isotropic, bringing the effective mass parameters closer to each other. It can be seen that the two functionals incorporating J yield a band-structure that is almost identical to that of DFT+U(all atoms), indicating that the J corrections have little effect on the rutile band-structure, except for the gap. These latter variations, while interesting, are difficult to base assessment upon, given that experimental estimations of the electron effective mass in rutile can vary by an order of magnitude[74,75].

### 3.3 Monoclinic $ZrO_2$ results

The two O atoms in the $ZrO_2$ formula unit are subject to two different chemical environments, with one O atom being 3-fold coordinated and another being 4-fold coordinated. As a result, different U and J values can be calculated and applied to each of the O atoms[27]. Table 4 shows the convergence of calculated U and J parameters for $ZrO_2$ for Zr and the two different O environments. The values from the largest supercell are used in this study.

**Table 4: Convergence of calculated U and J parameters for $ZrO_2$ for different supercell sizes and k-point grids. The U and J values were calculated separately for the Zr 4d subspace, 3-fold coordinated O atom 2p subspaces, and 4-fold coordinated O atom 2p subspaces.**

| Supercell size | k-points | Zr U (eV) | Zr J (eV) | O 3-fold U (eV) | O 3-fold J (eV) | O 4-fold U (eV) | O4-fold J (eV) |
|---|---|---|---|---|---|---|---|
| unit cell | 3x3x3 | 1.724 | 0.346 | 14.126 | 2.342 | 15.665 | 2.564 |
| 2x2x2 | Γ-point | 1.735 | 0.337 | 14.277 | 2.317 | 15.590 | 2.559 |
| 2x2x2 | 2x2x2 | 1.740 | 0.336 | 14.250 | 2.327 | 15.590 | 2.532 |
| 3x3x3 | Γ-point | 1.736 | 0.338 | 14.277 | 2.339 | 15.453 | 2.564 |

Fig. 4a. shows the effect of U and J on the crystallographic geometry error. The PBE volumetric error is increased slightly when DFT+U is applied to Zr 4d subspaces alone, but when a U value is applied to the O atoms as well, the volumetric error decreases well below the PBE value, due to an underestimation of the b-axis size cancelling out slight overestimations of the a and c axes. In the case of U and J, the b axis length and $\beta$ angle match almost exactly with experimental values, but the a and c axis lengths are larger, leading to a volumetric error that is about the same as in uncorrected PBE simulations. This indicates that, aside from the most commonly used DFT+U approach of targeting Zr 4d orbitals only, no methodology distorts the ionic geometry significantly.

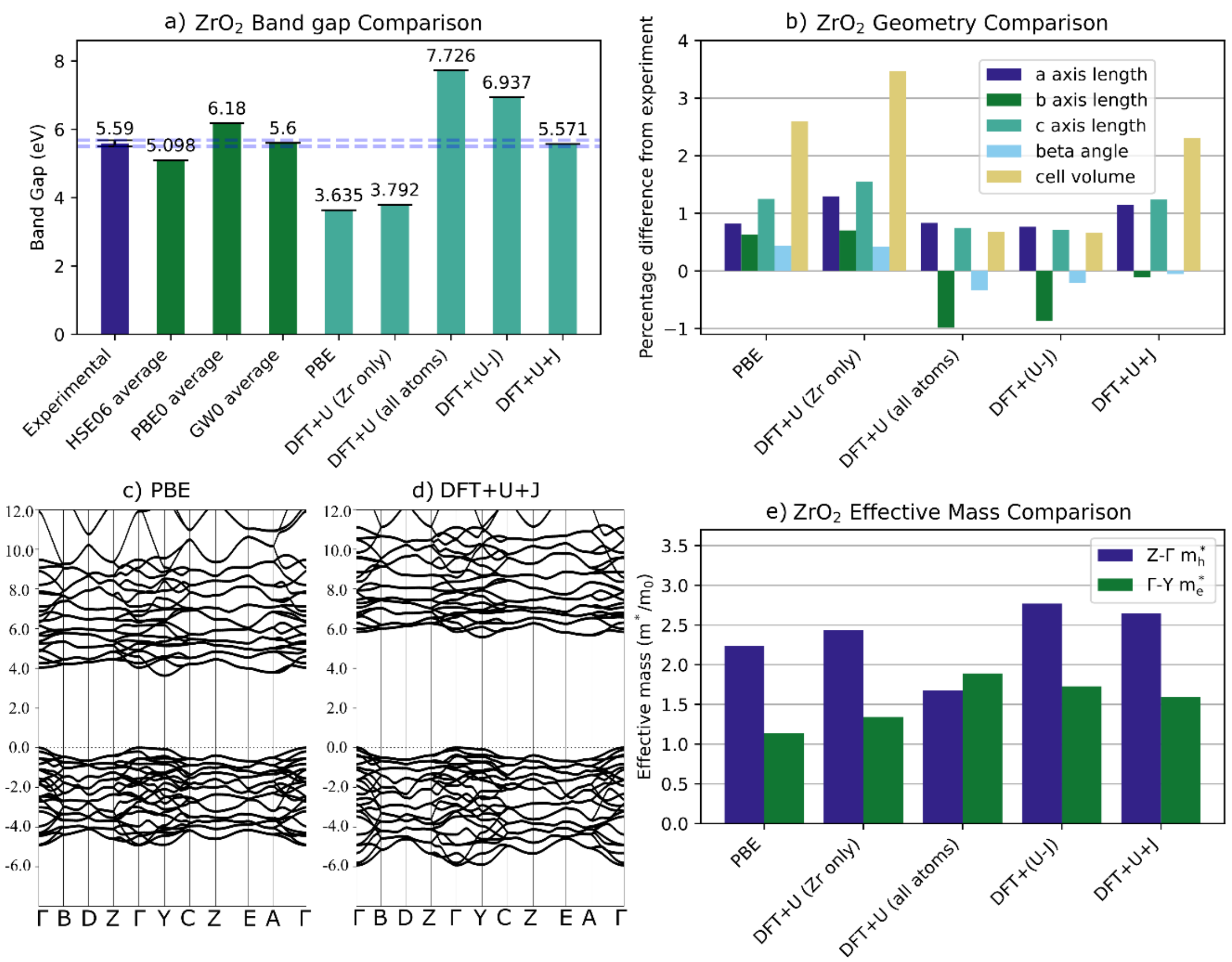

**Fig. 4. Summary of effect of different U and J incorporating corrective functionals on $ZrO_2$, showing: a) The effect of methodology on band-gap, compared to experiment and the literature values for HSE06[76-79], PBE0[80,81] and $GW_0$[82-84]. b) The effect of methodology on unit cell geometries, as a percentage difference from the experimental values[85,86]. c) The band-structure for PBE. d) The band-structure for DFT+U+J e) Comparison of calculated effective mass values along selected directions.**

Fig. 4c. and Fig. 4d. show the band-structures calculated using the PBE and DFT+U+J functionals, which are qualitatively very similar apart from the differing bandgaps. The effective mass of the electrons and holes are slightly increased when U and J are applied, but the results are not significantly different from PBE values. As with rutile $TiO_2$, there is an expansion of the valence band width, from 4.9 eV for PBE to 5.9 eV DFT+U+J.

Overall, the first-principles DFT+U+J approach seems to be highly efficient at correcting the band-gap to the experimental value without distorting geometry or band-structure. This negates the pre-supposition that DFT+U methods are fundamentally inapplicable to $d^0$ or $d^{10}$ systems. We emphasize that both Hund's coupling J and oxygen 2p terms are needed for satisfactory results, for different reasons.

### 3.4 Monoclinic $HfO_2$ results

Monoclinic $HfO_2$ is similar to $ZrO_2$ in that it comprises 3-fold and 4-fold coordinated O atoms that yield different calculated U and J values. Table 5 shows the convergence behavior of these parameters. The U and J values for the largest supercell are used.

**Table 5: Convergence of calculated U and J parameters for $HfO_2$ for different supercell sizes and k-point grid parameters. The U and J values were calculated separately for Hf 5d subspaces, 3-fold coordinated O atom 2p subspaces, and 4-fold coordinated O atom 2p subspaces.**

| Supercell size | k-points | Hf U (eV) | Hf J (eV) | O U 4-fold (eV) | O J 4-fold (eV) | O U 3-fold (eV) | O J 3-fold (eV) |
|---|---|---|---|---|---|---|---|
| unit cell | 3x3x3 | 1.442 | 0.327 | 17.875 | 3.139 | 16.071 | 2.863 |
| 2x2x2 | Γ-point | 1.443 | 0.321 | 17.755 | 3.092 | 16.350 | 2.811 |
| 2x2x2 | 2x2x2 | 1.438 | 0.324 | 18.065 | 3.115 | 16.251 | 2.819 |
| 3x3x3 | Γ-point | 1.441 | 0.323 | 17.815 | 3.144 | 16.187 | 2.829 |

Similarly to the case of $ZrO_2$, the experimental gap of $HfO_2$ is not well known. We again use an average of IPES studies to estimate the fundamental bandgap as 5.78 eV[85,86]. Fig 5a. shows that $G_0W_0$ calculations in the literature have yielded values that match within 0.01 eV of this value, with $GW_0$ results slightly overestimating it by 0.28 eV. As is the case with $TiO_2$ and $ZrO_2$, the PBE0 functional overestimates the band-gap (by 0.73 eV), while the HSE06 result is much closer (within 0.09 eV). In our simulations, the PBE functional underestimated the band-gap by 1.58 eV, with the underestimation only improving to 1.46 eV when the DFT+U correction was applied to Hf 5d subspaces alone. Applying U to both O 2p subspaces results in a very large bandgap overestimation of 3.37 eV, which is slightly reduced to 2.42 eV by the Dudarev DFT+(U-J) functional. The PBE+U+J approach once again yields the most accurate bandgap of the cost-effective PBE-derived approaches, with an overestimation of 0.61 eV, which is more accurate than PBE0 for this system but falls short of the HSE06 performance.

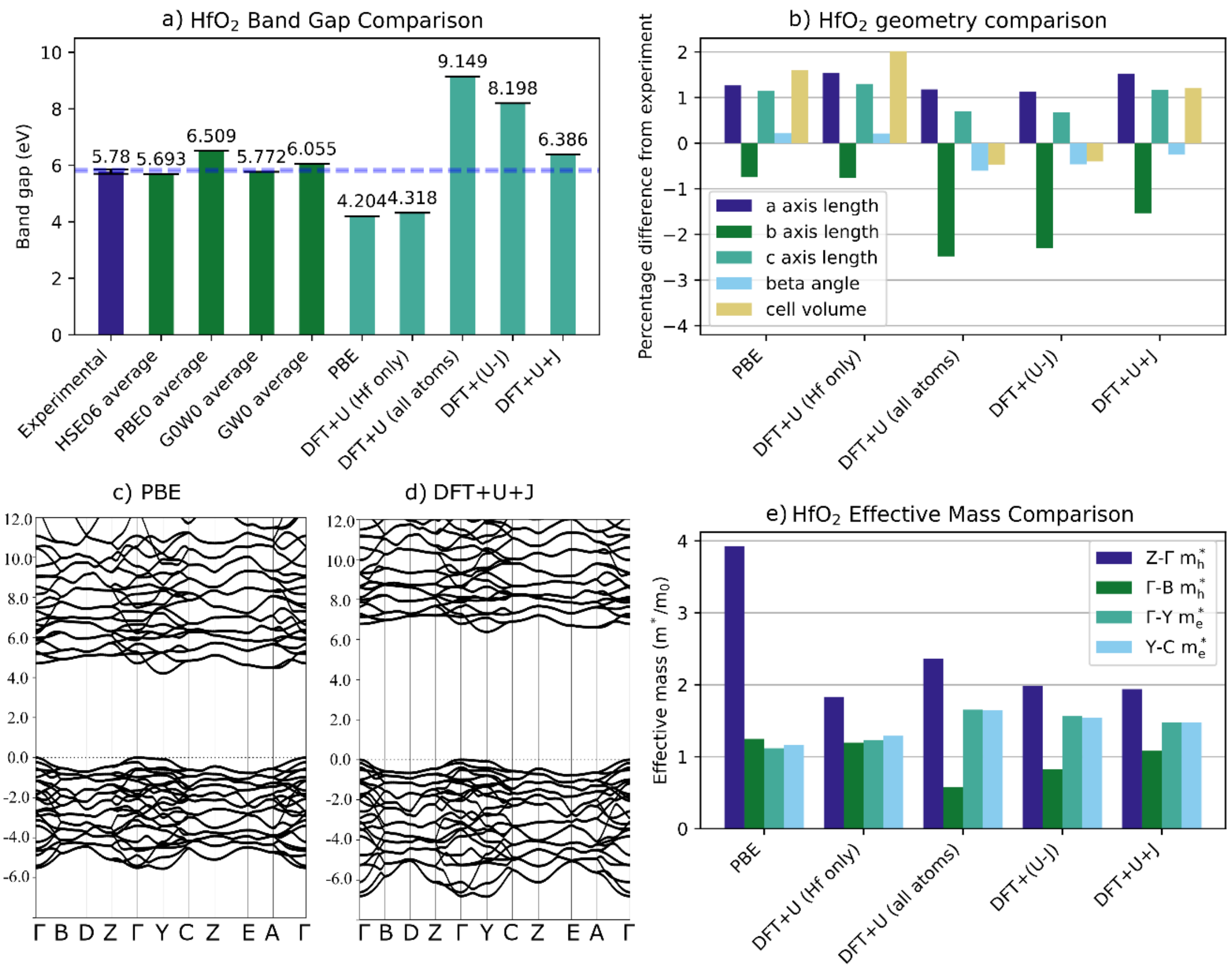

**Fig. 5. Summary of effect of different U and J incorporating corrective functionals on $HfO_2$ material properties, showing: a) Calculated bandgaps from different methodologies, compared to experiment[85,86], HSE06[87-89], PBE0[87,90,91], $G_0W_0$[83,92,93], and $GW_0$[82,83,92]. b) Difference with respect to experiment values for crystallographic cell properties and total volume for each method. c) Band-structure from uncorrected PBE. d) Band-structure from the first-principles DFT+U+J method. e) Effective mass of electrons and holes along selected directions for each functional.**

The geometric effect of the various U and J methods on $HfO_2$ is shown in Fig 5b. The application of U has a small stretching effect on the a and c axes, resulting in a slight increase in volumetric distortion. Applying the U additionally to the O atoms rectifies this somewhat, and also results in a large degree of shrinkage in the b-axis, resulting in a low volumetric error. The application of J reduces this shrinkage but increases the a-axis and b-axis lengths, resulting in a volumetric error that is ultimately slightly lower than the case of uncorrected PBE. Fig 5c. and Fig 5d. shows the band-structure of uncorrected PBE and DFT+U+J based on PBE, showing that there is not a large amount of change between them. Fig 5e. shows the resulting effect on effective mass. The application of the corrective functionals tends to increase the effective mass of electrons slightly. The hole effective mass is reduced, conversely, but since the lines are very flat it is difficult to assess the difference. As is the case for $TiO_2$ and $ZrO_2$, the valence band width is increased, now from 5.5 eV for PBE to 6.8 eV for DFT+U+J.

### 3.5 Cubic $Cu_2O$ results

The convergence of calculated U and J parameters is shown in Table 6, indicating a reasonable degree of convergence with the accessible supercell sizes. The band-gap of $Cu_2O$ was found by an IPES study to be 2.17[94]. Fig. 6a. shows the band-gaps of different methods compared to this result. Hybrid PBE0 modelling has overestimated the band-gap by 0.30 eV, while HSE06 models have slightly underestimated the bandgap by 0.24 eV.

**Table 6: Convergence of calculated U and J parameters for Cu and O atoms in cubic $Cu_2O$ with different supercell sizes and k-point sampling.**

| Supercell size | k-points | Cu U (eV) | Cu J (eV) | O U (eV) | O J (eV) |
|---|---|---|---|---|---|
| 2x2x2 | Γ-point | 12.382 | 1.850 | 20.358 | 3.199 |
| 2x2x2 | 3x3x3 | 12.450 | 1.961 | 20.801 | 3.183 |
| 3x3x3 | Γ-point | 12.525 | 1.933 | 20.601 | 3.168 |
| 3x3x3 | 2x2x2 | 12.537 | 1.960 | 20.551 | 3.188 |
| 4x4x4 | Γ-point | 12.526 | 1.958 | 20.240 | 3.203 |
| 5x5x5 | Γ-point | 12.476 | 1.954 | 20.407 | 3.191 |

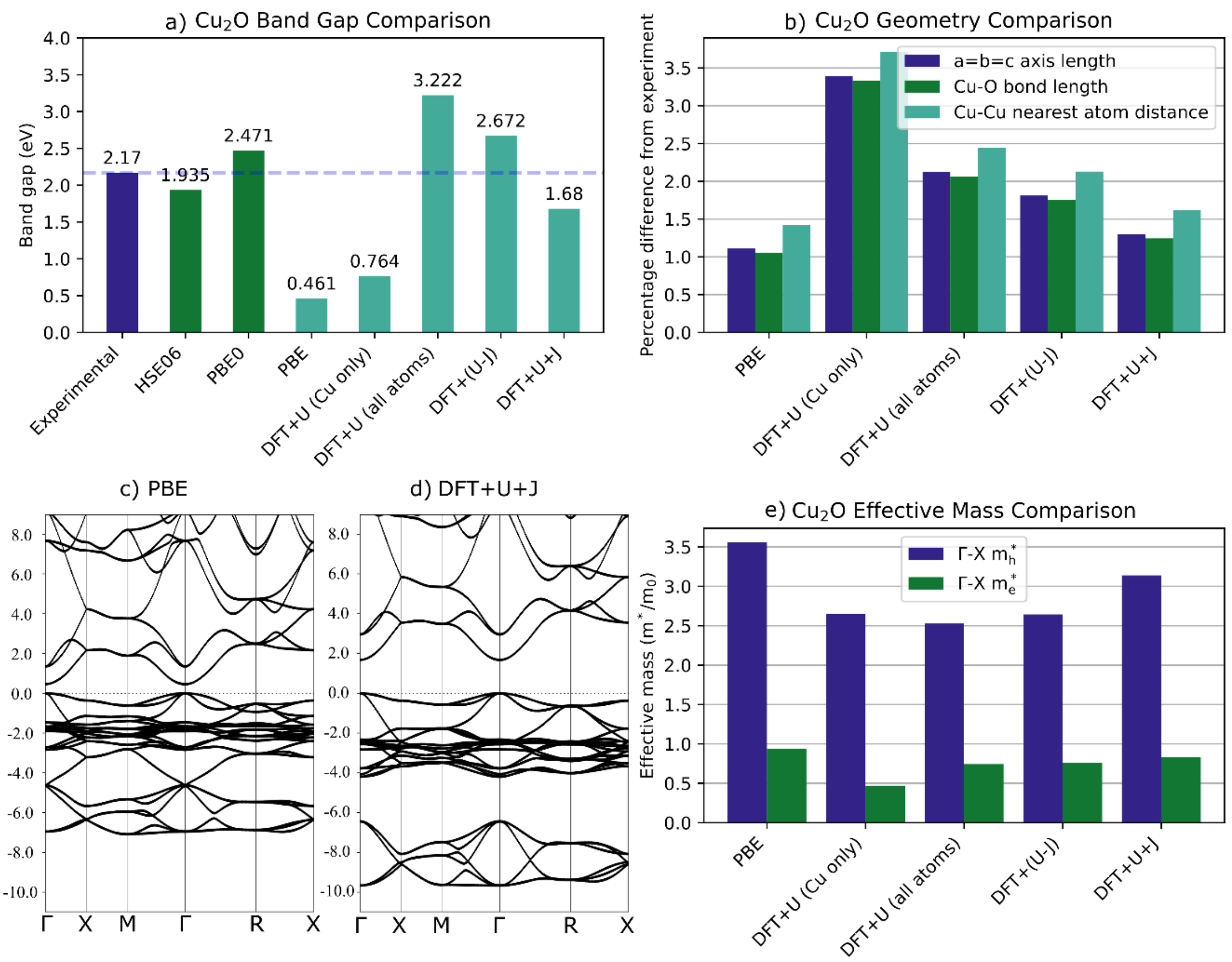

**Fig. 6. Summary of effect of different U and J incorporating corrective functionals on $Cu_2O$ material properties. a) Comparison of predicted bandgaps for the five functionals with the experimental value,[94] as well as an average of HSE06 literature values[17,95-99] and PBE0 values[17,98,100]. b) Percentage deviation of crystallographic properties (axis and inter-atomic distances) from experimental values for each method. c) PBE band-structure with no corrections applied. d) PBE structure with both U and J corrections applied to Cu and O. e) Effective mass in selected directions for the five functionals.**

Our calculated band-gap with PBE and no corrections greatly underestimates the band-gap by 1.71 eV. Applying the first-principles U value to the Cu 3d orbitals alone only lessens the overestimation to 1.41 eV. Applying the U to both the Cu and O atoms again results in an overestimation of the band-gap value, by 1.05 eV, which can be reduced by DFT+(U-J) to 0.50 eV. This is very similar to the final error of 0.49 eV for the DFT+U+J method, which instead errs on the side of underestimation. DFT+U+J is again the best performing of the PBE-based functionals but falls short of the hybrid accuracy in this case. The reason that DFT+U+J appears to be less accurate for $Cu_2O$ than it is for other materials is unclear, yet may lie in its distinctive bond character. It seemed, for this material, appropriate to investigate whether the use of ortho-normalized projectors could make a more suitable choice.[27] This was tested with a full recalculation of the results with ortho-atomic projectors, detailed in Appendix III. The DFT+U+J bandgap results were ultimately found to be less accurate with these projectors, although the DFT+U(all atoms) approach was found to yield a slight improvement. It would be worth investigating alternate projector profiles for Cu 3d orbitals, or even the inclusion of Cu 4s corrections, in future research. We emphasise, on a cautionary note, that $Cu_2O$ seems to represent a point close to the boundary of the applicability of DFT+U methods based on d-orbital projectors. The following material, ZnO, is beyond that boundary, albeit with a remedy available, as we go on to discuss.

The effect of corrective functional on crystallographic parameters is shown in Fig. 6b. The PBE geometry is quite accurate, with values that are within 1.5% of experimental values. Applying the DFT+U correction to the Cu 3d subspaces alone introduces some distortions, with cell parameter errors rising to about 3.5%. These distortions decrease back down to around 2% for DFT+U (all atoms) and DFT+(U-J). The DFT+U+J method reduces the errors still further, and turns out to be almost as good as uncorrected PBE. This indicates that DFT+U+J does not significantly distort the geometry of $Cu_2O$ that is already predicted well by PBE.

The band-structure of DFT+U+J is qualitatively quite similar to the PBE one near the CBM and valence band maximum (VBM), as can be seen by a comparison of Fig. 6c. and Fig. 6d. However, the DFT+U+J functional does result in the opening of a ‘second gap’ in the valence band that is not present in the PBE band-structure, which starts 4.2 eV below the VBM and has a gap width of 2.2 eV. Such second gaps can be of interest for photovoltaic and other optoelectronic functionalities[101,102]. A previous quasiparticle $G_0W_0$ study[17] shows a dip to zero or near-zero DOS at around the level predicted here for the second gap, but it is significantly less wide (<0.5 eV) than that found in this study. PES results do not seem to readily support the existence of such a gap[103]. This suggests that the second gap found using DFT+U+J may be either overestimated in magnitude, or the gap finding may be entirely erroneous. Ref. [104] indicates that the hybrid functional HSE06 does not predict this feature in $Cu_2O$. The effective masses, shown in Fig. 6e, do not change substantially with the various tested functionals. The largest deviation occurs for DFT+U applied to metal only, where the $m^*_e$ is about half of the PBE value.

### 3.6 Wurtzite ZnO results

Previous attempts at applying the linear response methodology to calculate the Hubbard U for Zn 3d orbitals in ZnO have encountered great difficulties, with researchers encountering either excessively high calculated U parameters[23,28] or numerical instability[29,30]. We have found that the almost perfectly fully filled Zn 3d orbitals require the applied perturbation to be increased by an order of magnitude in order to produce a sufficient change in occupancy level to overcome the numerical noise of the simulation. However, this unsurprisingly results in a response that is nonlinear. U and J values for these nonlinear responses can, in principle, still be estimated by taking the slope at zero perturbation of a good parabolic fit to the response curve. We found the resulting U and J values to be very high, with a 5x5x5 supercell calculation yielding the remarkable values of 83.7 eV for U and 9.3 eV for J. A DFT+U+J bandgap of 6.35 eV was calculated with these values, reflecting a saturating effect on the band-gap when the corrections are applied to a subset of orbitals. This is more than 3eV higher than the experimental gap, indicating that a simple strategy of always applying U to the metal d-orbitals will sometimes fail, if said orbitals are near-fully occupied. This could be deemed a pit-fall of the methodology, but it is one that is already evident of the very hard and ultimately non-linear response of the Zn 3d orbital subspace. Put simply, if the linear response calculation does not go smoothly, we recommend that the choice of orbital, or its projector profile, be reconsidered.

It has been shown that DFT+U based modelling can be improved by applying U corrections to the 4s orbitals of certain transition metal systems.[29] This motivated us to calculate the linear response parameters of the Zn 4s subspaces, which is more relevant to the character of the band edges than the Zn 3d ones, particularly at the conduction band edge. For this it was necessary to change the PWscf source code trivially on a few lines, so that the desired angular momentum was selected. The charge response of the partially filled 4s subspace is much better behaved, and linear response can be readily extracted for the same perturbation range as for the four prior materials, yielding U and J values that are much more plausible, as shown in Table 7. These values (for the largest supercell) were used for our evaluation.

**Table 7: Convergence of calculated U and J parameters for Zn 4s and O 2p subspaces in cubic ZnO with different supercell sizes and k-point grids.**

| Supercell size | k-points | Zn U (eV) | Zn J (eV) | O U (eV) | O J (eV) |
|---|---|---|---|---|---|
| unit cell | 6x6x6 | 1.753 | 0.953 | 19.293 | 3.910 |
| 2x2x2 | Γ-point | 1.404 | 0.824 | 19.641 | 3.296 |
| 2x2x2 | 2x2x2 | 1.765 | 0.988 | 23.189 | 3.911 |
| 3x3x3 | Γ-point | 1.743 | 0.993 | 22.540 | 3.871 |
| 3x3x3 | 2x2x2 | 1.795 | 1.072 | 23.428 | 3.954 |
| 4x4x4 | Γ-point | 1.820 | 1.092 | 23.325 | 3.906 |
| 5x5x5 | Γ-point | 1.815 | 1.066 | 23.325 | 3.906 |

Fig. 7a. shows the bandgap comparison for ZnO. The experimental bandgap is 3.44 eV[105]. Hybrid functionals consistently underestimate this gap, with the HSE06 average being 0.97 eV below experiment and PBE0 having a smaller error of 0.25 eV. Bare PBE drastically underestimates the gap, by 2.70 eV. Interestingly, applying the DFT+U correction to the Zn 4s orbitals alone actually makes the band-gap smaller, with the resulting error increasing to 3.11 eV. This is rectified by applying corrections to the O atoms as well, and the DFT+U (all atoms) method produces a band-gap that actually outperforms HSE06 as assessed from the literature, with an overcorrection of 0.63 eV. Both methods of applying J improve this slightly, with the 0.4 eV error of DFT+(U-J) making it the best performing of the PBE-based methods in this instance. The DFT+U+J functional gives a very close result to this, however, with an overestimation of 0.42 eV.

Since the effect of O 2p correction on the bandgap is significantly greater than the effect of the Zn 4s correction, it is worth checking whether the Zn correction is necessary *at all* for accurate results. Two more bandgap calculations were therefore performed for ZnO with U and U+J applied to the O 2p orbitals only and not to the metals atoms, yielding bandgaps of 4.27 eV and 2.95, respectively. This is slightly less accurate than the DFT+U+J on both metal and O, with an underestimate by 0.49 eV instead of the 0.42 eV overestimate of the full methodology, indicating that the application of U on Zn is valuable.

Fig. 7b. shows the effect of U correction on crystallographic geometry. The Zn 4s correction alone has almost no effect on the geometry. Applying U to the O atoms using either the DFT+U (all atoms) or DFT+(U-J) methods results in a shrinking of axis and bond lengths that overcorrects PBE somewhat. This still gives a lower geometric error overall than uncorrected PBE. The DFT+U+J method increase the error slightly, but it is still lower than that of PBE. This indicates that neither the unconventional choice of Zn 4s orbitals to define the subspaces targeted for correction, nor the somewhat high-seeming calculated O 2p U values, result in ionic geometry distortion. In fact, the geometry is improved with respect to that of PBE.

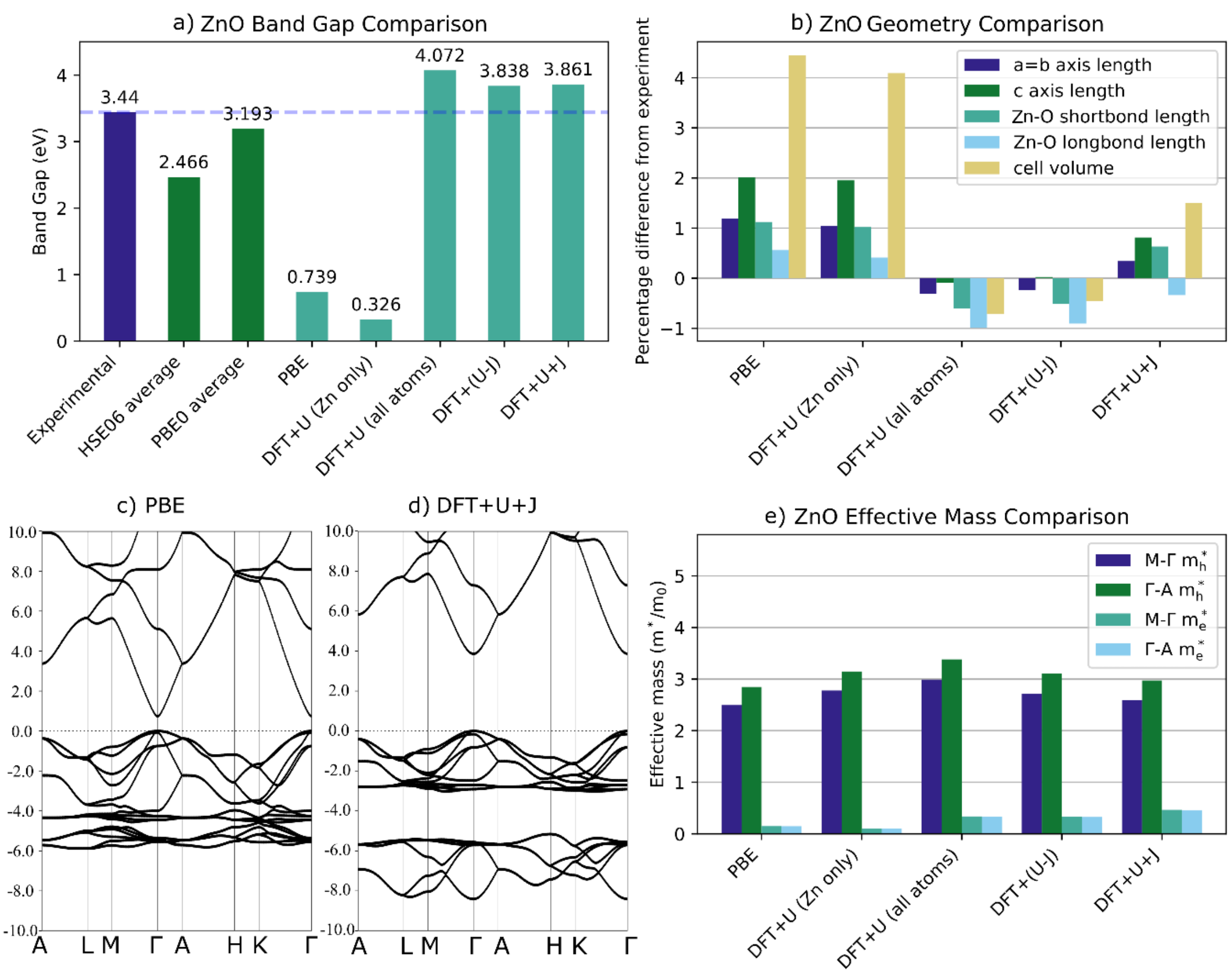

**Fig. 7. Full comparison of methodology choice effect on ZnO properties with corrections applied to 4s orbitals of Zn and 2p orbitals of O, showing the a) effect on band-gaps compared to experiment,[106] HSE06,[105,107-110] and PBE0[105,108,110].**

**b) geometric property differences with respect to experimental values. c) band-structure for PBE. d) band-structure for DFT+U+J. e) effective mass of along selected directions.**

A comparison of Fig. 7c. and Fig. 7d. indicates that the introduction of U and J introduces a second gap within the valence band that is not present in the bare PBE calculation, which starts at 3.0 eV below the VBM and has a gap width of 2.1 eV. This opening of this second gap has also been observed in some other DFT+U studies[111,112], some quasiparticle GW calculations by Kotani *et al.*[113] and in XPS data[114], although in general these second gaps appear to be much less wide than is seen here (<1eV). In contrast, most uncorrected GGA band-structures do not have a second gap[112,115], and the second gap similarly does not appear to be present in some HSE06 studies[105,107]. The effective mass of holes remains largely unchanged between methods, as can be seen in Fig. 7e. The effective mass of electrons, on the other hand, is very substantially increased multiplicatively.

### 3.7 Trends across the material test set

The calculated U and J for each orbital of each material have been collated in Table 8.

**Table 8: Summary of calculated U and J values for each material considered in this work.**

| Material | Atom | Orbital | U (eV) | J (eV) | U:J ratio |
|---|---|---|---|---|---|
| $TiO_2$ | Ti | 3d | 3.24 | 0.38 | 8.43 |
| $ZrO_2$ | Zr | 4d | 1.74 | 0.34 | 5.14 |
| $HfO_2$ | Hf | 5d | 1.44 | 0.32 | 4.46 |
| $Cu_2O$ | Cu | 3d | 12.48 | 1.95 | 6.39 |
| ZnO | Zn | 4s | 1.81 | 1.07 | 1.70 |
| $TiO_2$ | O | 2p | 11.24 | 1.70 | 6.62 |
| $ZrO_2$ | O (3-fold) | 2p | 14.28 | 2.34 | 6.10 |
| $ZrO_2$ | O (4-fold) | 2p | 15.45 | 2.56 | 6.03 |
| $HfO_2$ | O (3-fold) | 2p | 16.19 | 2.83 | 5.72 |
| $HfO_2$ | O (4-fold) | 2p | 17.81 | 3.14 | 5.67 |
| $Cu_2O$ | O | 2p | 20.41 | 3.19 | 6.40 |
| ZnO | O | 2p | 23.32 | 3.91 | 5.97 |

There are several trends of note across the material set for the calculated U and J values, with the oxygen results all being high compared to every metal, with the exception of Cu. These may be partially explained by the spatial extent of this orbital. Fig. 8 shows a graph of the radial density of each orbital for an isolated atom, which appears to have some qualitative relation to the U and J value. For example, the Zn 4s, Zr 4d, and Hf 5d orbitals have quite similar radial density profiles, and are fairly diffuse, resulting in similarly low U values. In contrast, the Cu 3d and O 2p orbitals are much more short ranged, and have higher U values. We observe a possible (not statistically significant) relationship whereby the U is proportional to the inverse square of the spread (central second moment) of the metal projector orbitals, when we limit the analysis to the four metal atoms with larger metal-centred projector orbitals (in $TiO_2$, $ZrO_2$, $HfO_2$, and ZnO). Only in these four metals is the spread calculated based on the non-ultrasoft-augmented density expected to be reliable. This also does not explain the full extent of the variations, as the same O 2p orbitals have U values varying by a factor of two across the four oxides. It can be expected that other factors such as orbital filling, orbital coordination, dielectric screening and details of the approximate functional being measured will also play a role in the U value.

The relatively high values for the O2p orbital are consistent with the results from other linear response studies, for example linear response calculations of the O 2p U in rutile $TiO_2$ have yielded values in the range from 8.6 to 15.9.[27,37,116] The value will be highly dependent on the exact methodology used, for example Ref. [27] calculated a U value for 4-fold O 2p in $HfO_2$ of 47.8 eV for atomic projectors, but only 9.58 eV for ortho-atomic projector calculations.[27]. Another study using cRPA calculated screened interaction U for O 2p-like Wannier functions in TMO's ranging from 3 to 7 eV, depending on the interaction model used.[117] It is interesting to ask why the calculated Hubbard U parameter, or put another way the localized many-body self-interaction error in the underlying approximate functional, is relatively high in O 2p subspaces. In a very recent work at the time of writing, in Ref. [32], it is argued the periodic-table trends in U parameters may be partially explainable, as a direct result of the formula used for U, in terms of trends in the chemical hardness of the isolated atoms in question. This quantity, which is related to energy-occupancy curvature by definition, broadly increases as we move up and right in the periodic table, and its value is particularly large for oxygen. Moreover, in systems such as $TiO_2$ where the valence band is very heavily dominated by the O 2p orbitals, it is argued that only the ionization energy component of the chemical hardness matters, and this quantity is again particularly large for atomic oxygen, relative to that of the transition metals in question here.

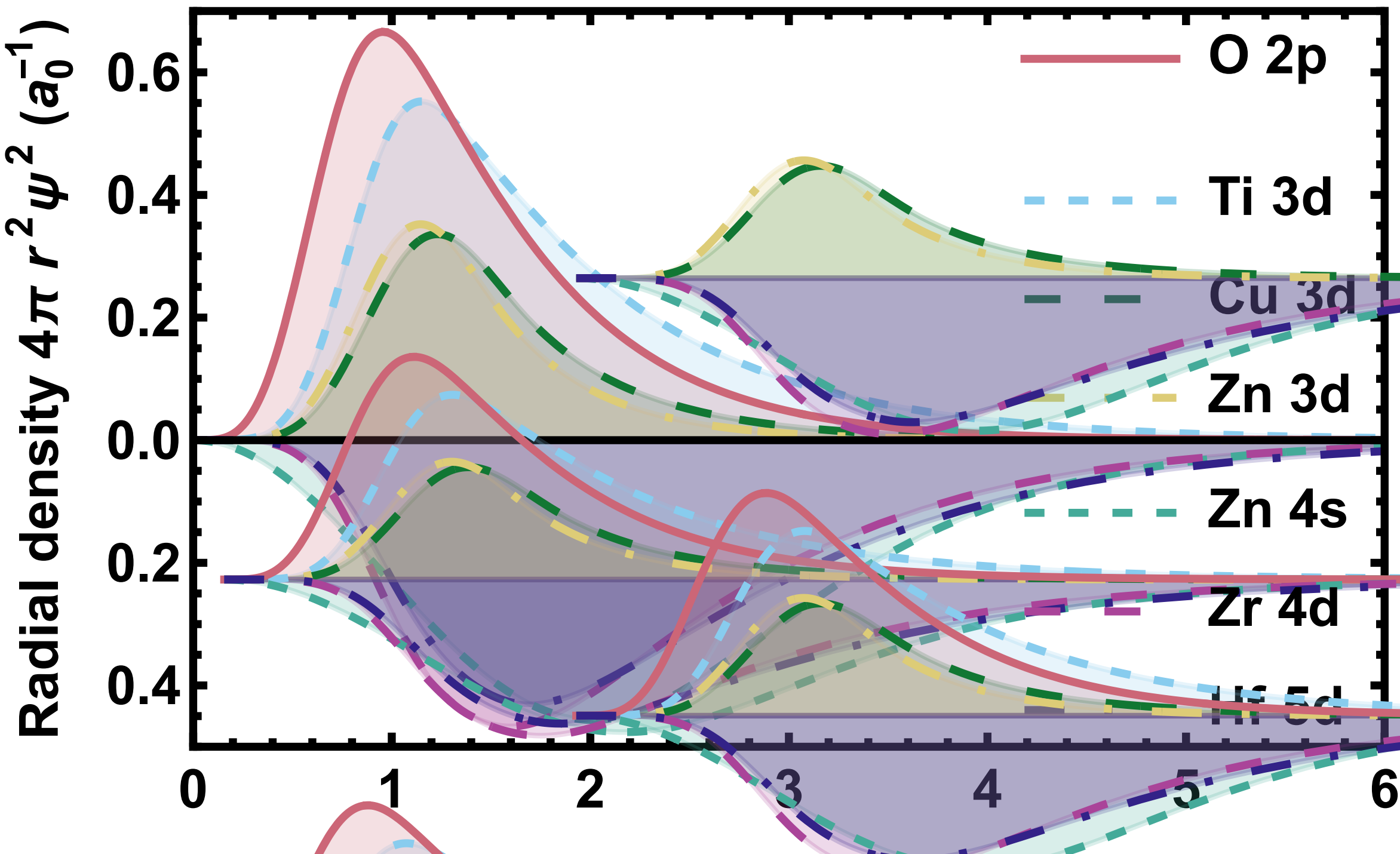

**Fig. 8: Radial density distribution (atomic units) of the pseudo-atomic orbitals used to define the Hubbard projectors for corrective functionals in this study. The projector-augmented wave core-region modifications are not shown. Beyond 1.5 Bohr, the effective all-electron Hubbard projectors are as shown for all species.**

Finally, we note that the U:J ratios found for the O 2p orbitals are in a similar range for each material, varying by only 10% from the average value, despite the U values varying by a factor of two. Similarly consistent ratios can be found in the work of Chai et al[116], with their ratio ranging from 9.58 to 9.87 for four different oxides. This ratio does not seem to be transferable to other orbitals, indeed the U:J ratios in the metal atoms vary significantly. We emphasize that within the present formalism, these computed values and ratios therefore reflect more about pathologies in the spin-polarised extension of the PBE functional, specifically energy curvatures of different types, than anything directly about Coulomb or exchange interactions per se.

Table 9 summarizes the effect of the various corrective functionals on the band-gap for the materials investigated (For a summary of the actual values used for this calculation, see appendix II). Uncorrected PBE underestimates the bandgap of all materials by a significant degree, ranging from 27% for $HfO_2$ to 79% for ZnO. Applying DFT+U to the metal atoms alone increases the bandgap only slightly for each material except for ZnO (with 4s orbital subspaces targeted), where it actually makes the bandgap less accurate.

**Table 9: Evaluation of theoretical band-gap accuracy across five chemically diverse non-spin-polarized transition-metal oxides, using a variety of first-principles functionals derived from PBE and incorporating corrective Hubbard U and Hund's exchange coupling J parameters, alongside HSE06 and PBE0 hybrid functional values based on available literature.**

| | Band-gap error with respect to experiment (%) | | | | | |
|---|---|---|---|---|---|---|
| Functional | $TiO_2$ | $ZrO_2$ | $HfO_2$ | $Cu_2O$ | ZnO | MAE |
| PBE | -39 | -35 | -27 | -79 | -79 | 52 |
| DFT+U (metal only) | -33 | -32 | -25 | -65 | -90 | 49 |
| DFT+U (all atoms) | 26 | 38 | 58 | 48 | 18 | 38 |
| DFT+(U-J) | 13 | 24 | 42 | 23 | 12 | 23 |
| DFT+U+J | -8 | 0 | 10 | -23 | 13 | 11 |
| HSE06 average | 7 | -9 | -2 | -11 | -28 | 11 |
| PBE0 average | 30 | 11 | 13 | 14 | -7 | 15 |

When the U is applied to both the metal and O atoms, an overestimation of the bandgap occurs in all five materials, with the largest by percentage occurring in $HfO_2$. Using the DFT+(U-J) functional to scale down the effective U improves this slightly, but this still overestimates the gap for all materials.

The most comprehensive technique tested, that of applying both U and explicit unlike-spin J corrections in DFT+U+J, gives rise to a moderate overestimation of the bandgap for the 5d metal-oxide $HfO_2$ and the arguably marginal transition-metal oxide ZnO, a similar underestimation of the bandgap for the 3d metal-oxide $Cu_2O$ and $TiO_2$, and the correct value within the experimental uncertainty for the 4d metal $ZrO_2$. At least for the titanium Group 4 metal-oxides, the 4d pseudoatomic orbitals for PBE thus appear to represent a 'goldilocks zone' where the population analysis happens to be well suited to the assumptions underpinning DFT+U type methods. This conclusion might not be transferable to other periodic table groups, or other underlying functionals, we hasten to emphasize.

Meanwhile, the literature indicates that HSE06 slightly underestimates the gap on a mean-absolute error (MAE) basis, and that PBE0 overestimates it. The MAE of first-principles DFT+U+J is 11%, which is lower than that of all of the other PBE-derived functionals, lower than that of PBE0 from literature, and equal to that of HSE06 values from the literature. This demonstrates the DFT+U+J can provide band-gap accuracy comparable to hybrid functionals, and typically at a small fraction of the computational cost after the parameters are computed.

Table 10 shows a similar comparison of the effect of U and J on the crystallographic unit cell volume of each material. It can be seen that DFT+U applied to only the metal increases the volumetric error for every material except ZnO, and that is has the highest error of all the methods investigated. However, once the U correction is applied to the O atoms on the same footing as the metal atoms, the volumetric error drops to below that of regular PBE. This remains the case when J is applied using either of the functionals tested. The best performing functional for the lattice volume within this test set is DFT+(U-J), but the full DFT+U+J functional still provides a lower error than PBE. This demonstrates that accurate bandgaps do not need to come at the expense of spurious lattice distortion, as has previously been found in several studies[18-20], the key evidently being to correct the ligand band-edges subspaces (oxygen 2p in this study) on the same footing as the metal ones.

**Table 10: Comparison of accuracy of simulated volume across all five materials for each methodology of applying U and J corrections.**

| | Unit cell volume error with respect to experiment (%) | | | | | |
|---|---|---|---|---|---|---|
| Functional | $TiO_2$ | $ZrO_2$ | $HfO_2$ | $Cu_2O$ | ZnO | MAE |
| PBE | 2.81 | 2.60 | 1.60 | 3.37 | 4.45 | 2.96 |
| DFT+U (metal only) | 4.60 | 3.47 | 2.02 | 10.52 | 4.11 | 4.94 |
| DFT+U (all atoms) | 0.77 | 0.68 | -0.47 | 6.50 | -0.72 | 1.83 |
| DFT+(U-J) | 0.96 | 0.66 | -0.40 | 5.54 | -0.42 | 1.59 |
| DFT+U+J | 2.59 | 2.30 | 1.21 | 3.95 | 1.60 | 2.33 |

Table 11 shows the difference in valence bandwidth between uncorrected PBE and DFT+U+J. The DFT+U+J functional, for the materials and subspace choices selected, consistently either stretches the valence bandwidth, or split it into two separate bands, predicting a "second gap" in CuO and ZnO. The appearance of a second gap in the valence band is a qualitative distinction that seems to feature in available many-body perturbation theory calculations for the $Cu_2O$ and ZnO band-structures, yet seems to be predicted by few if any DFT simulations except those of DFT+U or self-interaction correction type[17,105,106,111,112].

**Table 11: Comparison of valence bandwidths calculated using PBE and first-principles DFT+U+J constructed on the basis of PBE and PBE pseudoatomic orbitals.**

| | Valence bandwidth (eV) | | | | | | |
|---|---|---|---|---|---|---|---|
| Technique | $TiO_2$ | $ZrO_2$ | $HfO_2$ | $Cu_2O$ 1st band | $Cu_2O$ 2nd band | ZnO 1st band | ZnO 2nd band |
| PBE | 5.63 | 4.91 | 5.50 | 7.08 | n/a | 5.88 | n/a |
| DFT+U+J | 6.31 | 5.92 | 6.83 | 4.19 | 3.23 | 3.03 | 3.25 |

# 4. Conclusion

In this study the use of linear response DFT+U+J was investigated as a means to accurately model closed-shell metal oxide bandgaps using unmodified, widely available plane-wave DFT software. It was demonstrated that Hund's coupling J parameters can be calculated successfully from first-principles using the familiar SCF linear-response formalism. The "γ-method" was extended and verified within the SCF linear-response formalism, which allows for simultaneously calculating U and J in ultimately non-spin-polarized systems. This essentially makes the J a cost-free by-product of a U calculation.

Several corrective functionals incorporating Hubbard U and Hund's J corrections were evaluated in detail on a chemically diverse test set made up of $TiO_2$, $ZrO_2$, $HfO_2$, $Cu_2O$ and ZnO. The fundamental bandgap, crystallographic geometry, and carrier effective masses were highlighted. It was shown that DFT+U+J (applied to both metal and O orbitals on the same footing) was overall the most successful functional for modelling the bandgap among those tested. First-principles DFT+U+J yielded the same mean-absolute band-gap error (MAE) as the popular, but much more computationally demanding and poorly-scaling hybrid functional HSE06.

Our results indicate that the magnitude of the derivative discontinuity introduced by DFT+U+J, that is U-2J for the non-spin-polarized and idealized case of the same J at both band edges and a further correction otherwise, seems to be sufficient to open the band-gap with a comparable level of reliability to popular hybrid functionals, without the costly introduction of the long-ranged part of the exchange. It is important to emphasize, however, that the choice of projector orbitals remains arbitrary in this method, so that this correct magnitude in practice may be the result of a fortunate error cancellation. In particular, given the relatively large first-principles U and J values for certain orbitals including O 2p, the dependence on the orbital profile and degree of orthonormalization may be important. It remains a matter for future studies to explore further the impact of different projector choices in DFT+U+J. Nonetheless, our data shows that following the default, physically motivated choices in the code used in the present work, PWscf, yields a very practical and relatively reliable first-principles approach for pragmatic users who wish to study oxides in large supercells and avoid the use of costly hybrid functionals.

Successful, or at least state-of-the-art level band-gap prediction, does not come at the cost of spurious crystallographic geometry distortion, as has been found in many previous DFT+U studies. The key difference in the present work being that the errors within the oxygen 2p subspaces were measured and corrected on the same footing as in the metal nd ones. In fact, the cell volumes predicted by DFT+U+J simulations were more accurate than those of standard PBE. Similarly, the effect on effective mass and band-structure tended to be small, although the bandwidth of the valence band was generally increased and a splitting of the valence band into two sub-bands was observed in the late transition-metal oxides $Cu_2O$ and ZnO.

The success of this technique on this particular test set is promising, and it implies that the technique should work well for similar materials in the range of non-magnetic TMOs. The success of this technique on other dielectric materials, such as magnetic TMOs, non-transition metal oxides, and nitrides, are topics of ongoing and future research. The relatively poor performance of $Cu_2O$ and of ZnO (when U is applied to 3d orbitals) suggests that mixed orbital characters may be a particular challenge for this technique.

Overall, DFT+U+J was found to be a highly viable method, which can be used easily within the widely available package Quantum Espresso. Moreover, it can be easily introduced into any DFT+U code, and no software modification at all is necessary if restricting its application to ultimately non-spin-polarized systems. We anticipate that the more widespread adoption of Hund's coupling J comprising corrective functionals, as a means to strengthen exchange effects locally at very low computational cost, could significantly improve the reliability of future materials discovery projects, while minimizing their environmental and financial footprint. Future research in this area could, e.g., examine the applicability of the methodology presented in this work to spin-polarized oxides, in which case Eqs. (9,10), not the γ-method, are required.

# Acknowledgements

This publication has emanated from research supported in part by a research grant from Science Foundation Ireland (SFI) and is co-funded under the European Regional Development Fund under Prime Award Number 12/RC/2278_P2. We also acknowledge Trinity College Dublin Research IT and the DJEI/DES/SFI/HEA Irish Centre for High-End Computing (ICHEC) for the provision of computational facilities and support. Calculations were performed using the Kay cluster at ICHEC and using three different clusters maintained by the Trinity Centre for High Performance Computing, the clusters Lonsdale, Kelvin and Boyle being funded by grants, respectively, from SFI, The Higher Education Authority through its PRTLI program, and the European Research Council and SFI.

We are pleased to acknowledge financial support from Intel Corporation and insightful discussions and guidance from Justin Weber and Karthik Krishnaswamy. DDO'R wishes to thank and acknowledge Edward B. Linscott, Christoph Wolf, Carsten A. Ullrich, and Okan K. Orhan for prior discussions that encouraged a detailed study into calculating Hund's J using a self-consistent field DFT code.

# Appendix I: Sign and pre-factor convention for Hund's J

In this appendix, we briefly explore the definition used for Hund's J in this work, in terms of a simple two-spin model. This serves to demonstrate how the convention used for quantifying spin-density in DFT determines the sign and pre-factor definition of Hund's J, without the rather involved (compatible) analysis of Ref. [36]. The spin-polarization, sometimes called magnetization in collinear spin DFT is defined, for the total within a subspace, is defined as $M = n^{\uparrow} - n^{\downarrow}$, as mentioned in the main text. Thus, spin here is measured in electrons, with magnitude of 1, not in terms of the spin angular momentum. If we consider the spin-interaction between two electrons with no external field, an appropriate model in terms of sign and magnitude convention is therefore the classical Heisenberg model

$$\hat{H} = -J\,\vec{S_1}\cdot\vec{S_2} \tag{24}$$

With $|\vec{S_\iota}| = 1$. This is the same convention used for Hund's J in DFT+U family methods.

Still considering this two-electron system, and envisaging dissociated $H_2$, if this is studied with an approximate density-functional then in general this description will suffer from static correlation error. This means that the energy $E_S$ of the singlet state of the system (spins anti-aligned) will be higher in energy than that of the triplet state (spins aligned), $E_T$, where these energies would be equal in the absence of static correlation error and externally applied magnetic field. Mapping this onto the aforementioned Heisenberg model, we have $E_S - E_T = 2J$. Permitting the magnetization to vary continuously and assuming a quadratic static correlation error that matches this result, we may write

$$E(M) = E_S - \frac{J}{2}M^2 \tag{25}$$

By examination, if we evaluate this at the fully aligned state, we obtain

$$E(M = 2) = E_S - 2J = E_T \tag{26}$$

as expected. Taking the second derivative of this energy model with respect to magnetization, we recover the basic definition of Hund's J for measuring global static correlation error in approximate density functional theory, $J = -\frac{d^2E}{dM^2}$.

In practice, in linear-response methods for calculating such parameters, one focuses on derivatives of the potential (on a subspace), rather than second derivatives of the energy. The magnetization is perturbed using a subspace uniform potential of value + $\beta$ for spin-up and - $\beta$ for spin-down. This is exactly the potential choice that minimally increases the total energy while perturbing the magnetization, as it is consistent with the constrained DFT total energy,

$$W = E_{DFT} + \beta(M - M_{target}) \tag{27}$$

where $\beta$ plays the role of a Lagrange multiplier[58]. Supposing that the cDFT stationary point is identified always, we may consider variations with respect to the desired magnetization $M_{target}$, which is the external parameter. We note that M will always equal $M_{target}$ at the cDFT solution. First recalling the definition of J, and then using the Hellmann-Feynman theorem, we find that

$$-\frac{d^2E}{dM_{target}^2} = \frac{\partial\beta}{\partial M_{target}} = +\frac{1}{2}\,\frac{\partial(v_{ext}^{\uparrow} - v_{ext}^{\uparrow})}{\partial M_{target}} \tag{28}$$

since beta is half of the difference between the spin-up and spin-down perturbation strengths. This is nothing but the 'total' part of J, making up half of the terms in Eq. (8) of the main text. Restricting J to only measure the interacting part of the curvature, since it will be used only in a functional that directly modifies interaction (not kinetic energy, explicitly), we can then subtract the non-interacting part to recover the full Eq. (8) of the main text, which simplifies to Eq. (7), which serves as the core definition of J. In conclusion, the definition of the spin-density difference M in terms of electron count, rather than electron angular momentum, taken together with Ising's sign convention that J is positive for aligned moments, determines the sign and multiplicative pre-factor for calculating Hund's J from first-principles using linear-response DFT, that is Eq. (9) for self-consistent field linear-response DFT.

# Appendix II: Full bandgap results:

Table 12 summarizes the absolute bandgap values for all considered materials, used for the percentage error calculations in

Table 9. The values for the additional method of applying U and J to just the metal orbitals and not the O orbitals are included here. The latter method did not produce an accurate bandgap for any material, and for $TiO_2$, $ZrO_2$ and $HfO_2$ it was almost identical to the value for an application of only U to the metal. In the case of ZnO, the bandgap was slightly improved, but it was still less accurate than even bare PBE.

**Table 12: Summary of bandgaps values for all materials in this study.**

| | Material Bandgap (eV) | | | | |
|---|---|---|---|---|---|
| Method | $TiO_2$ | $ZrO_2$ | $HfO_2$ | $Cu_2O$ | ZnO |
| Experimental | 3.03 | 5.59 | 5.78 | 2.17 | 3.44 |
| HSE06 average | 3.25 | 5.10 | 5.69 | 1.94 | 2.47 |
| PBE0 average | 3.94 | 6.18 | 6.51 | 2.47 | 3.19 |
| PBE | 1.83 | 3.64 | 4.20 | 0.46 | 0.74 |
| DFT+U (metal only) | 2.04 | 3.79 | 4.32 | 0.76 | 0.33 |
| **DFT+U+J (metal only)** | **1.98** | **3.77** | **4.32** | **0.66** | **0.61** |
| DFT+U (All atoms) | 3.83 | 7.73 | 9.15 | 3.22 | 4.07 |
| DFT+(U-J) | 3.43 | 6.94 | 8.20 | 2.67 | 3.84 |
| DFT+U+J | 2.79 | 5.57 | 6.39 | 1.68 | 3.89 |

# Appendix III: Ortho-atomic projector test

A test was made of the use of ortho-atomic projectors for $Cu_2O$, with U and J recalculated for a 3x3x3 supercell with 2x2x2 k-points. The resulting U and J values for Cu are 9.87 eV and 1.39 eV, and the U and J values for O 2p orbitals are 9.01 eV and 1.25 eV.

The results for bandgap are shown Fig. 9a Unlike in the default projector case, here the DFT+U (all atoms) approach underestimates the bandgap, and this underestimate increases when J is applied. It should be noted that the DFT+U (all atoms) approach slightly outperforms our DFT+U+J approach with default projectors, warranting further investigation of ortho-atomic projectors in future research.

Fig. 9b shows a comparison for lattice geometry. Similarly to the case with default projectors in Fig. 6, the highest distortion occurs when the U is applied to the Cu alone, but this is improved to levels similar to the bare PBE when U and J are applied to all atoms.

For the sake of completeness, the bandstructure and effective mass were also compared in Fig. 9c-e. The main difference between these results and that of default projectors shown in Fig. 6 is that there no longer appears to be a second bandgap in the valence band.

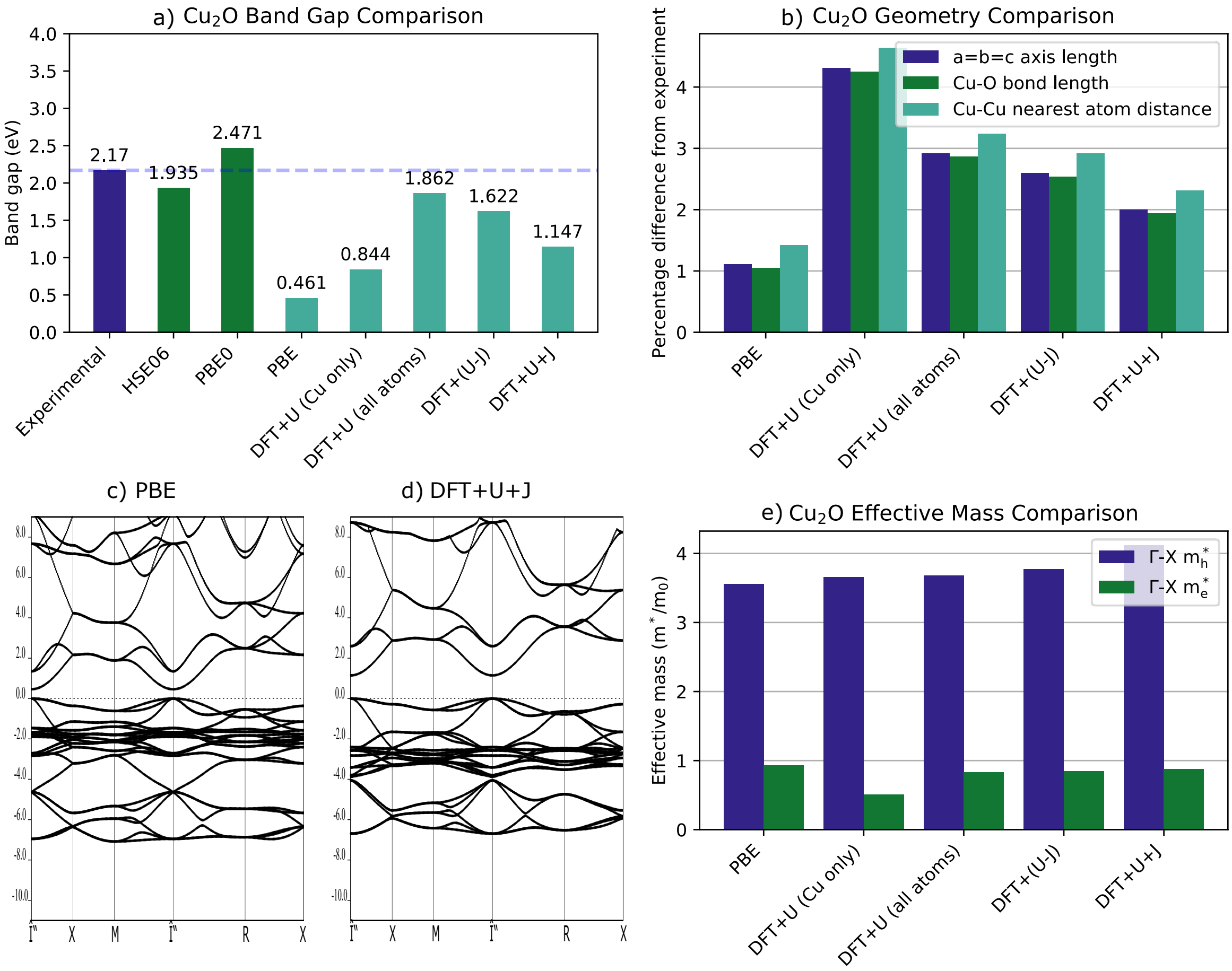

**Fig. 9: Summary of effect of different U and J incorporating corrective functionals on $Cu_2O$ material properties, implemented using ortho-atomic projectors. a) Comparison of predicted bandgaps for the five functionals with the experimental value,[94] as well as an average of HSE06 literature values[17,95-99] and PBE0 values[17,98,100]. b) Percentage deviation of crystallographic properties (axis and inter-atomic distances) from experimental values for each method. c) PBE band-structure with no corrections applied. d) PBE structure with both U and J corrections applied to Cu and O. e) Effective mass in selected directions for the five functionals.**

# Appendix IV: Results tables

Table 2 in section 3.1 demonstrated that the γ-method introduced in section 2.3 gave the same results as separate *α*-for-U and *β*-for-J self-consistent field linear-response calculations for $TiO_2$. Table 13 below summarizes similar tests for the other four materials. In all cases the difference between calculated values is very small, indicating that the γ-method is accurate for each material.

**Table 13 Demonstration that the U and J values calculated for $ZrO_2$, $HfO_2$, $Cu_2O$ and ZnO with separate *α* and *β* linear response calculations and with the γ-method enabling simultaneous U and J calculation, are near-identical.**

| Material | Parameter | α and β method (eV) | γ-method (eV) | Difference (eV) | Difference (%) |
|---|---|---|---|---|---|
| ZrO2 | Zr U | 1.728 | 1.724 | -3.943E-03 | -0.228 |
| | Zr J | 0.344 | 0.346 | 1.806E-03 | 0.525 |
| | O U | 14.187 | 14.126 | -6.082E-02 | -0.429 |
| | O J | 2.340 | 2.342 | 1.832E-03 | 0.078 |
| HfO2 | Hf U | 1.440 | 1.442 | 1.496E-03 | 0.104 |
| | Hf J | 0.328 | 0.327 | -1.225E-03 | -0.373 |
| | O U | 17.939 | 17.875 | -6.469E-02 | -0.361 |
| | O J | 3.162 | 3.139 | -2.371E-02 | -0.750 |
| Cu2O | Cu U | 12.112 | 12.165 | 5.287E-02 | 0.436 |
| | Cu J | 1.978 | 1.980 | 2.095E-03 | 0.106 |
| | O U | 19.225 | 19.286 | 6.135E-02 | 0.319 |
| | O J | 3.259 | 3.250 | -8.425E-03 | -0.259 |
| ZnO | Zn U | 1.751 | 1.753 | 2.044E-03 | 0.117 |
| | Zn J | 0.978 | 0.953 | -2.478E-02 | -2.535 |
| | O U | 19.312 | 19.293 | -1.936E-02 | -0.100 |
| | O J | 4.006 | 3.910 | -9.503E-02 | -2.372 |